\documentclass[conference]{IEEEtran}
\IEEEoverridecommandlockouts
\usepackage{amssymb}
\usepackage{amsmath}
\usepackage{amsthm}
\usepackage{amsfonts}
\usepackage{mathrsfs}
\usepackage{float}
\usepackage{graphicx}
\usepackage{epsfig,graphics,subfigure,psfrag}
\usepackage{pdfpages}
\usepackage{enumerate}
\usepackage{engord}
\usepackage{cite}
\usepackage{verbatim}
\usepackage[ruled]{algorithm2e}
\begin{document}

\newtheorem{definition}{\bf Definition}
\newtheorem{remark}{\bf Remark}
\newtheorem{theorem}{\bf Theorem}
\newtheorem{lemma}{\bf Lamma}
\newtheorem{proposition}{\bf Proposition}
\newcommand{\titlefontsize}{\fontsize{19pt}{20}\selectfont}

\title{\titlefontsize Real-Time Fine-Grained Air Quality Sensing Networks \\ in Smart City: Design, Implementation and Optimization}

\author{\IEEEauthorblockN{Zhiwen Hu, Zixuan Bai, Kaigui Bian, Tao Wang, and Lingyang Song}
\IEEEauthorblockA{\textit{State Key Laboratory of Advanced Optical Communication Systems and Networks,}\\
\textit{School of Electronics Engineering and Computer Science, Peking University, Beijing, China}\\
\{zhiwen.hu, zixuan.bai, bkg, wangtao, lingyang.song\}@pku.edu.cn}
}

\author{
    {Zhiwen~Hu},~\IEEEmembership{Student~Member,~IEEE},
    {Zixuan~Bai},~\IEEEmembership{Student~Member,~IEEE},
    {Kaigui~Bian},~\IEEEmembership{Senior~Member,~IEEE},\\
    {Tao~Wang},~\IEEEmembership{Senior~Member,~IEEE},
    {and~Lingyang~Song},~\IEEEmembership{Fellow,~IEEE}
\thanks{Manuscript received September 14, 2018; revised December 17, 2018; accepted February 18, 2019.
This work was supported by the National Nature Science Foundation of China under grant number 61625101.
The associate editor coordinating the review of this paper and approving it for publication was Prof. Sherman Shen.}
\thanks{Z.~Hu, Z.~Bai, K.~Bian, and T.~Wang are with the National Engineering Laboratory
for Big Data Analysis and Applications, School of Electronics Engineering and Computer Science, Peking University, Beijing, China. L.~Song is with the State Key Laboratory of Advanced Optical Communication Systems and Networks, School of Electronics Engineering and Computer Science, Peking University, Beijing, China. (Email: \{zhiwen.hu, zixuan.bai, bkg, wangtao, lingyang.song\}@pku.edu.cn).}
\thanks{Copyright \copyright  2012 IEEE. Personal use of this material is permitted. However, permission to use this material for any other purposes must be obtained from the IEEE by sending a request to pubs-permissions@ieee.org.}
}

\maketitle

\thispagestyle{empty}
\begin{abstract}
Driven by the increasingly serious air pollution problem, the monitoring of air quality has gained much attention in both theoretical studies and practical implementations.
In this paper, we present the architecture, implementation and optimization of our own air quality sensing system, which provides real-time and fine-grained air quality map of the monitored area.
As the major component, the optimization problem of our system is studied in detail.
Our objective is to minimize the average joint error of the established real-time air quality map, which involves data inference for the unmeasured data values.
A deep Q-learning solution has been proposed for the power control problem to reasonably plan the sensing tasks of the power-limited sensing devices online.
A genetic algorithm has been designed for the location selection problem to efficiently find the suitable locations to deploy limited number of sensing devices.
The performance of the proposed solutions are evaluated by simulations, showing a significant performance gain when adopting both strategies.
\end{abstract}


\section{Introduction}

Based on a recent report of the World Health Organization~\cite{bib_WHO}, air pollution has been proved to be one of the greatest threat to human health, which is responsible for one in eight of deaths each year.
In addition to the exhaust emission from industrial production procedures, the daily activities of residents also contribute to the accumulation of air pollutants, such as driving fuel automobiles or incinerating garbages~\cite{bib_AriPollution}.
The degree of air pollution is usually quantitatively described by the air quality index (AQI), which is defined according to the concentrations of some typical air pollutants, including the fine Particulate Matters (e.g., PM$_{2.5}$ and PM$_{10}$) and other basic chemical substances~\cite{bib_AQI}.
The value of AQI will be larger if the concentrations of the air pollutants become higher, indicating a higher risk of people suffering from harmful health effects.

To measure the concentration of a specific air pollutant, available approaches could be either the large professional instruments with high precision or the tiny commercial sensors with low cost~\cite{bib_Mosaic}.
For the consideration of accuracy, the government-owned official meteorological bureaus have deployed authoritative monitoring systems across the country with high costs.
Despite the high precision they can achieve, these official systems only have limited numbers of observation stations over a large area and provide measurement results with significant latency~\cite{bib_Station}.
However, recent studies show that the concentrations of air pollutants have the intrinsic characteristics to change from meters to meters, especially for the particulate matters in the urban areas with complicated terrain resulted from densely distributed tall buildings~\cite{bib_Meter2Meter,bib_Vertical}.
This indicates that the data provided by official measurements lose their accuracy to represent the air quality at remote locations.

Therefore, it is preferred that large number of low-cost tiny sensing devices are deployed to provide air quality sensing for the regions with complicated terrain~\cite{bib_BlueAer,bib_UAir}.
Since the deployment of tiny sensing devices can be dense and the data collection can be frequent, the air quality distribution can be updated with low latency and high resolution~\cite{bib_YuzheIOT,bib_YuzheGLOBECOM}.
Such a solution creates a promising application of Internet-of-Things (IoT) in smart city~\cite{bib_IoTAir}, where massive data can be collected and analyzed~\cite{bib_IoTData}.
The citizens are able to benefit from the valuable information provided by the air quality sensing system, by following the suggestions like keeping away from the highly polluted area or deciding the best ventilation system for a building~\cite{bib_Suggestions}.

In this paper, we propose the architecture, implementation and optimization of our own air quality sensing system, which provides real-time and fine-grained air quality map of the monitored area.
For the system design, a four-layer architecture is established, including the energy-efficient sensing layer, the high-reliable transmission layer, the full-featured processing layer, and the user-friendly presentation layer.
For the implementation, we have deployed this system in Peking University (PKU) for six months and have collected over $100$ thousand data values from $30$ devices.
The terrain of our campus is considered complex enough to represent a typical urban terrain of a large smart city, since green areas, tall buildings and vehicle lanes are all included.
For the system optimization, we aim to minimize the error of the real-time and fine-grained air quality map, where the limited number of available sensing devices and the limited capacity of their batteries are the challenges.

As the major part of this paper, the optimization of the IoT air quality sensing network is studied in detail, which is rarely taken into account in related works~\cite{bib_Mosaic,bib_BlueAer,bib_AirCloud}.
Specifically, the necessity of performing optimization is essentially due to the fact that, the IoT air quality sensing devices are deployed without external power supply~\cite{bib_IoT,bib_Polluino} in order to adapt to the complicated measurement area.
Therefore, a sensing device can only perform a limited number of power-consuming actions, such as detecting the concentration of an air pollutant, or uploading data back to the server.
To recover a real-time and fine-grained air quality map from the sparse data, a procedure of inference and estimation is required, which can be realized by approaches such as machine learning~\cite{bib_Forecast,bib_Prediction}.
The accuracy of inferring the data at unmeasured locations and unmeasured times depends on the spatial-temporal structure of the collected data.
For instance, inferring the current air quality based on a measured value from long ago would be questionable~\cite{bib_YuzheINFOCOM}.
In addition, inferring the air quality at a certain location based on the data from a hardly correlated location is also inaccurate~\cite{bib_Infer,bib_YuzheICC}.
In order to guarantee the accuracy of the established air quality map, it is necessary to consider the problems of where to deploy the limited number of sensing devices (location selection problem) and when to perform sensing actions (power control problem).
These two problems are interdependent, e.g., the location selection could influence the correlation of the sensing data values and therefore influences the optimal power control.

In our work, we model the measurement error and the inference error based on the statistical data from our own system.
Our objective is to minimize the joint error of the real-time and fine-grained air quality map, by properly designing the power control and location selection strategies.
To be specific, the power control problem is solved by the proposed solution based on deep Q-learning by considering the system as a Markov Decision Process (MDP), which can be deployed online to deal with unexpected weather conditions.
The location selection problem is solved by the proposed genetic algorithm, which takes the result of $k$-means clustering as the initial genetic population and iteratively improves the location selection by widely searching the solution space.
Both solutions achieve satisfactory suboptimal outcomes, and the combination of our power control and location selection strategies presents a significant superiority to reduce the average joint error.
In addition, these solutions are scalable and therefore able to be implemented in a city-wide huge IoT air quality sensing network.

The main contributions of our work are listed as below:
\begin{enumerate}
\item We present our energy-efficient real-time and fine-grained air quality sensing system, which has been deployed in PKU for six months by Spet. 2018.
\item We model the measurement error and inference error in the air quality sensing system based on the collected data.
\item We provide a deep Q-learning solution for the power control problem to reasonably plan the sensing tasks of the power-limited sensing devices online.
\item We design a genetic algorithm for the location selection problem to efficiently find the suitable locations to deploy limited number of sensing devices.
\item The performance of the proposed solutions is evaluated by simulations, showing a significant performance gain when deploying both strategies.
\end{enumerate}

The rest of our paper is organized as follows.
Section II provides an overview of the design and implementation of our air quality sensing system.
Section III formulates the problem of minimizing the joint error.
Section IV discusses the parameters that influence the inference error.
Section V presents the deep Q-learning solution for power control.
Section VI presents the genetic solution for location selection.
Section VII shows the simulation results of the proposed solutions.
Finally, we conclude our paper in Section VIII.

\section{System Overview}\label{sec_Model}

In this section, we first provide a brief overview of the design of our air quality sensing system, and then present some of the representative implementation results, and finally describe the collected data set.

\begin{figure}[!thp]
\centering
\includegraphics[width=3.1in]{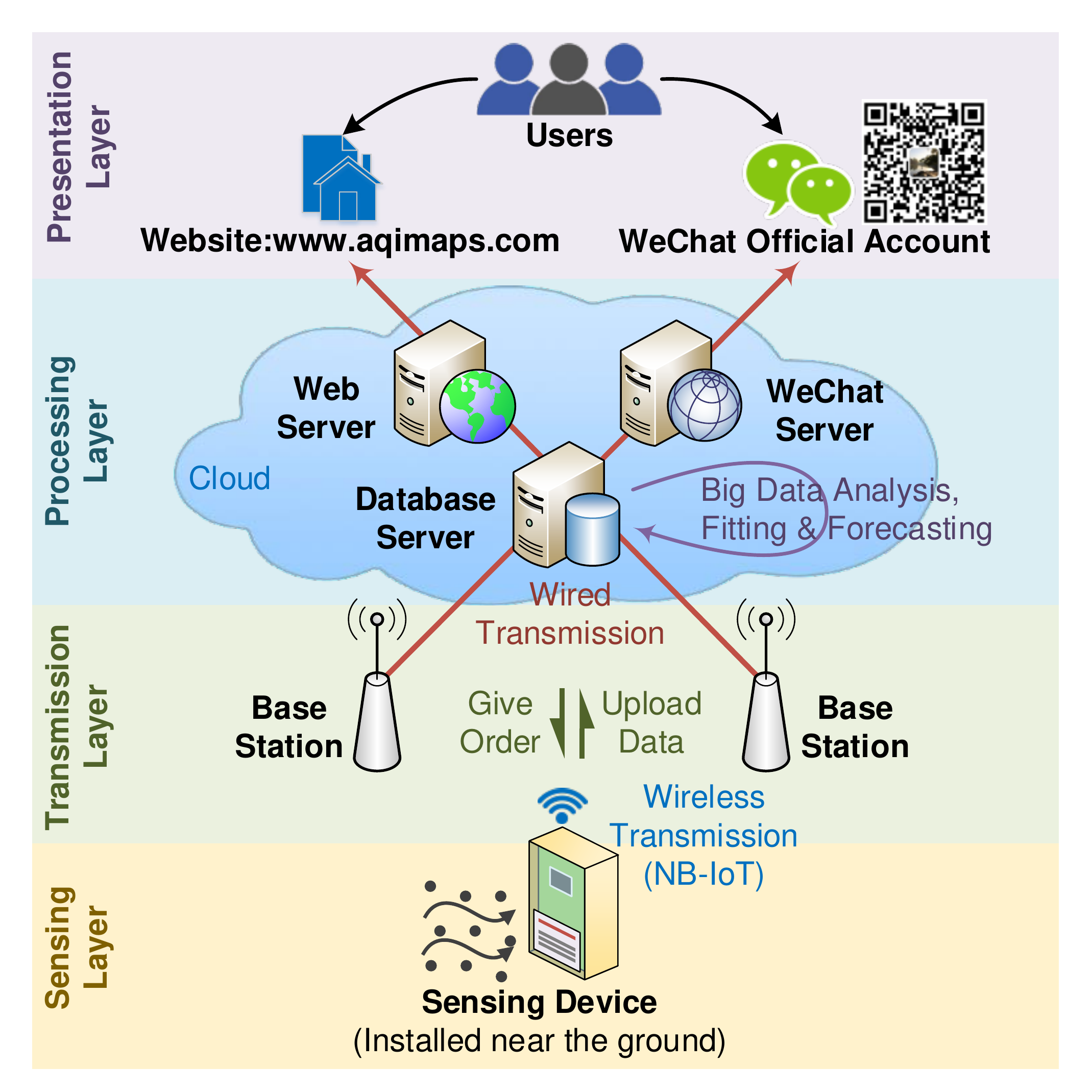}
\vspace{-3mm}
\caption{The architecture of our system, which consists of the sensing layer, the transmission layer, the processing layer and the presentation layer.}\label{fig_Architecture}
\vspace{-3mm}
\end{figure}

\subsection{System Design}

As shown in Fig.~\ref{fig_Architecture}, our air quality sensing system consists of four layers, namely, the sensing layer, the transmission layer, the processing layer and the presentation layer.
The sensing layer collects the data of real-time air quality, which is carried out by the sensing devices installed near the ground.
The transmission layer enables the bidirectional communications between the sensing layer and the processing layer, which is supported by the infrastructure of the current wireless communication networks.
The processing layer is implemented in the cloud server, which is responsible to receive, record and process the data from the sensing layer, and to control the behaviour of the sensing layer.
The presentation layer can provide valuable information for the users, which includes our official website and our official WeChat subscription account.

\begin{figure}[!thp]
\centering
\includegraphics[width=3.1in]{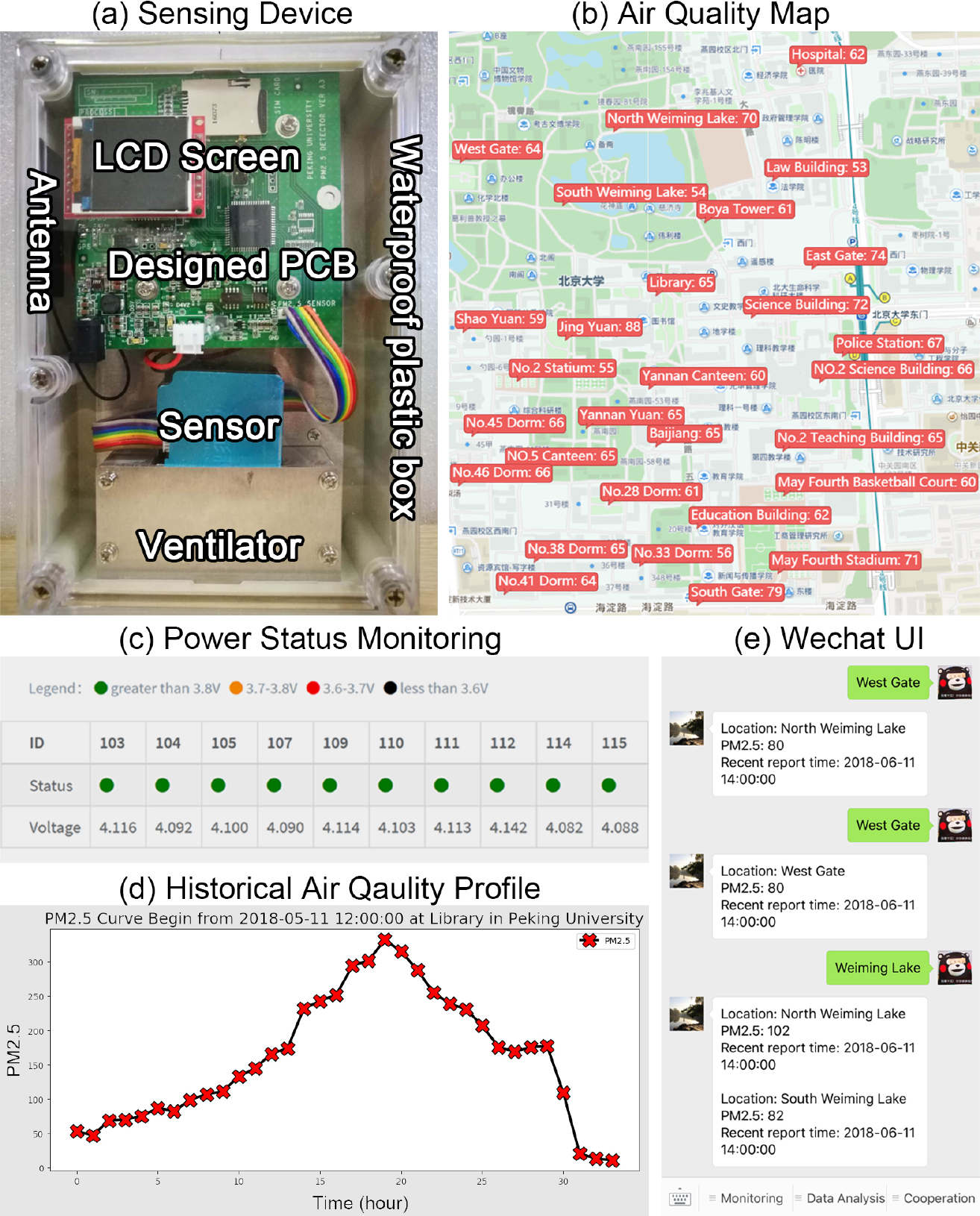}
\vspace{-3mm}
\caption{A simple exhibition of our system implementation.}\label{fig_Implementation}
\vspace{-3mm}
\end{figure}

\subsection{System Implementation}

Fig.~\ref{fig_Implementation} shows the implementation of our system, which has been deployed in PKU for $6$ months.
Most sensing devices are fixed near the ground and powered by batteries.
As the data being transmitted back to the server, users can inquire the real-time air quality data on our website~\cite{bib_Website} or through Wechat official account.
The backend of the server also monitors the status of the devices and manage their sensing behaviours to balance between accuracy and battery duration.
Spatial inference and short-term prediction can also be supported to guarantee the air quality map to be real-time and fine-grained.
More details can be found in~\cite{bib_Magazine}, which are not presented here, as we focus on the optimization of the air quality sensing network.

\subsection{Data Set Description}

During the deployment, we have collected over $100$ thousand effective values, mostly for the concentrations of PM2.5.
Here we provide the data set collected by $30$ on-ground sensing devices~\cite{bib_DataSet}.
Specifically, it contains the PM2.5 values from two time periods, including the period from March 1st 2018 to May 15th 2018, and the period from June 5th 2018 to Augest 25th 2018.
The provided data set is used to extract some important statistical properties of the monitored area, as given in Section~\ref{sec_Model}, in which way we are able to design the corresponding power control and location selection strategies.
If the proposed sensing system is expanded to the whole smart city, then the data set of the whole city will be necessary.

\section{Optimization Problem Formulation}\label{sec_Model}

In this section, we present the optimization problem in our air quality sensing system.
First, we provide the overview of the optimization problem in Section~\ref{sec_ModelOverview}.
We then model the measurement error and inference error in Section~\ref{sec_ModelError} based on the statistics of our collected data.
Finally, we formulate the optimization problem for the air quality sensing system, including power control and location selection.

\subsection{Problem Overview}\label{sec_ModelOverview}

The air quality sensor and wireless transmission module of each sensing device contribute to most of its power consumption.
Therefore, these devices keep themselves in sleep mode during most of the time to save their limited energy supplied by their own batteries.
The control server is responsible for planning the sensing tasks for all the devices (i.e., when should each device wake up and collect data), as well as receiving and recording the transmitted data.
Since the air quality data from nearby spatial locations and temporal points are not independent, the control server can utilize limited data to establish a real-time air quality map by spatial and temporal inference.

Assume that there are totally $K$ suitable locations for sensing deployment in the concerned area, and only $L$ sensing devices are available to be deployed, where $L<K$.
We denote the set of locations with sensing devices as $\mathcal{K}_L$, and the set of locations without sensing devices as $\mathcal{K}_U$.
Here, we have $\mathcal{K}_L\bigcup\mathcal{K}_U=\mathcal{K}$ and $\mathcal{K}_L\cap \mathcal{K}_U = \varnothing$.

The sensing system is divided into equal-length time slots, and we should decide whether each device is waken up to collect data at each time slot.
We provide the 0-1 matrix $\Phi_{K\times T}$ to represent the power control strategy, where $T$ is the expected number of time slots that the whole system should sustain without recharging.
As the element of the matrix $\Phi$, $\phi_{k\times t}=1$ indicates that the device in the $k^{th}$ location is turned on to sense data at the $t^{th}$ time slot, and $\phi_{k\times t}=0$ indicates that this device still keeps asleep or there is no device at the $k^{th}$ location.
The missing data values are inferred by the server, based on the current and previous collected data, according to their spatial-temporal relation.

\begin{figure}[!thp]
\centering
\includegraphics[width=3.1in]{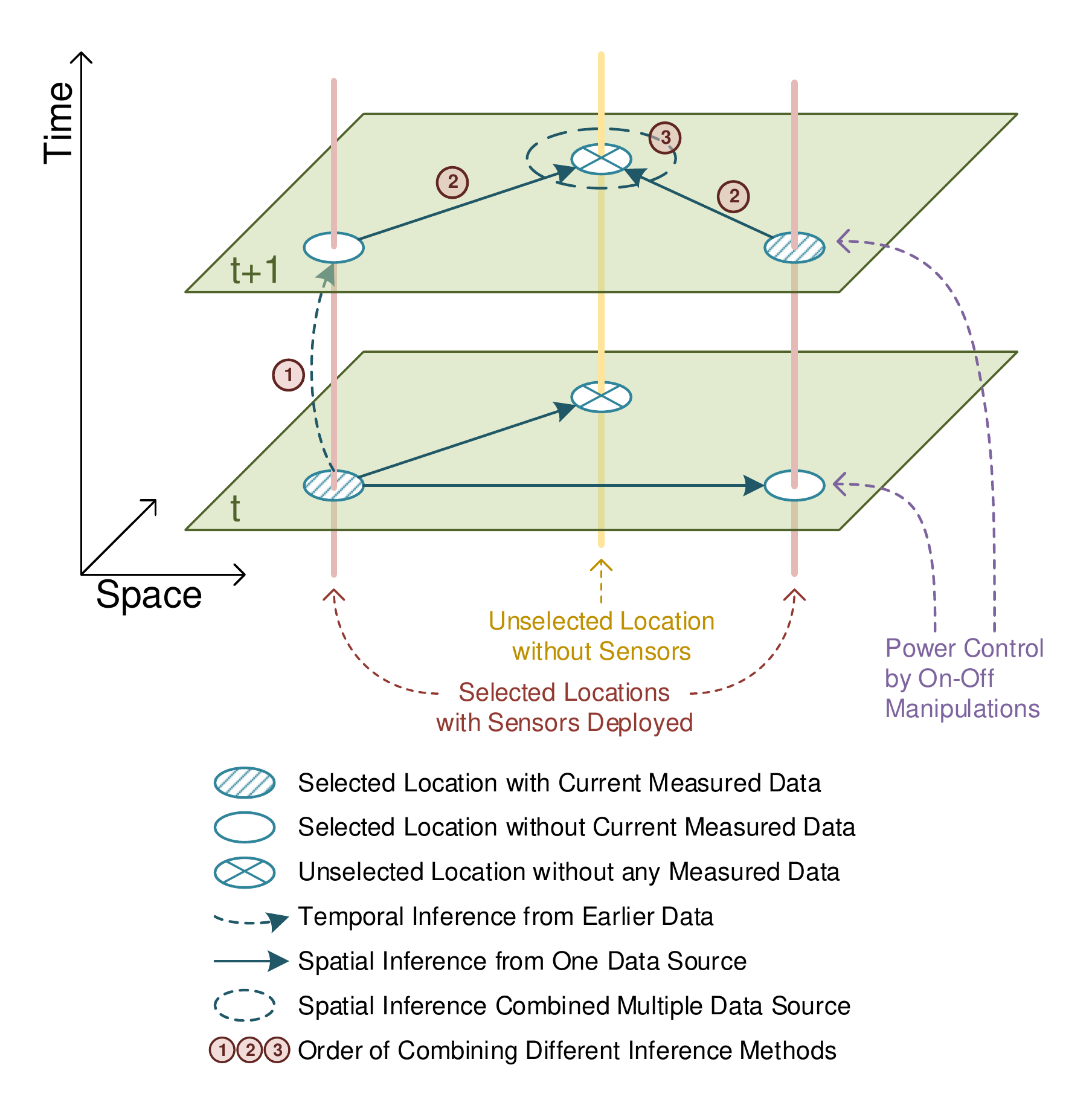}
\vspace{-3mm}
\caption{The spatial-temporal model of the air quality sensing system which contains spatial and temporal inference in order to provide real-time air quality map. Each inference results in different change in the distribution of the estimated data value.}\label{fig_SystemModel}
\vspace{-3mm}
\end{figure}

\subsection{Measurement Error and Inference Error}\label{sec_ModelError}

In this subsection, we model the measurement error and inference error based on our statistical data~\cite{bib_DataSet}.
The inference error here is modeled independently of any advanced inference algorithms that based on massive historical data (such as neural networks), in which way we can depict the most general situation.
The adopted inference method is based on Gaussian model, which accords with our collected long-term data as testified in Section~\ref{sec_SimulationResult}.
Due to its simplicity and analyzability, this inference method can be considered as a benchmark to depict the ``worst inference error" in the most general case.
By planning the location selection and power control strategy based on such an inference method, we are actually improving the worst-case performance of the system.

Regardless of whether the data is being directly measured or being inferred from other data, we denote the air quality value at the $k^{th}$ location at the $t^{th}$ time slot as a random variable $X_{k,t}$, where $k \in \mathcal{K}$ and $t=0,1,2,3,\cdots$.
In the following, the deviation of the mean value of $X_{k,t}$ and the uncertainty (variance) of $X_{k,t}$ are considered as the major indicators to represent the error of the measurement or the inference.

\textbf{Measurement:}
The measurements of the sensing devices are not perfect, the distribution of the measured value (e.g., PM2.5) at a ceratin location and a certain time approximately complies to Gaussian distribution\footnote{To be more precise, the values complies to truncated Gaussian distribution since the PM2.5 or any other air quality indicators should be $\ge 0$, but the small tail below zero can be ignored in most of the cases.}, given by
\begin{equation}\label{eqn_MeasureError}
\left\{\!\!\!\!
\begin{array}{ll}
 \!\!\!\!& X_{k,t} \sim \mathcal{N} (\mu_{k,t},\sigma_{k,t}^2), \\
 \!\!\!\!& \mu_{k,t} \approx \mu_t ,\\
 \!\!\!\!& \sigma_{k,t}^2 \approx \mu_{k,t}^2 \!\times\! \sigma_0^2,
\end{array}
\quad\forall t \ge 0, \, \forall k \in \mathcal{K},  \phi_{k,t}=1,
\right.
\end{equation}
where $\mu_{k,t}$ is the precise value of the $k^{th}$ location at the $t^{th}$ time slot\footnote{The precise PM2.5 value can be detected by a high-precision calibrating instrument TSI8530, which is expensive and not economical to be massively deployed.}, $\mu_t$ is the average value at the $t^{th}$ time, and $\sigma_0^2$ is a constant that reflects the common error of the adopted type of low-cost sensor.
And we call $\sigma_0^2$ as the normalized measurement variance.
We can see that the standard deviation $\sigma_{k,t}$ has a linear correlation with the precise value $\mu_{k,t}$ (or the average value $\mu_t$), which implies that the precision of the measurement decreases as the air quality is getting bad.

\textbf{Temporal inference:} With a measured value $X_{k,t}$ at location $k$ time $t$, we can infer the possible value at time $t+\tau$ for the same location.
As time goes on, the new value of this location deviates from the original one randomly.
Such deviation, can be seem as a additive random noise applied on the original measured value.
As long as the length of the time slot is fixed, the deviation between two adjacent time slots has a fixed distribution, given as
\begin{equation}\label{eqn_InferTimeError1}
X_d^{t\rightarrow t+1} \!\sim\! \mathcal{N} (0,\sigma_{d}^2), \quad\quad\quad\quad \forall t\ge 0,
\end{equation}
where $\sigma_{d}^2$ is the constant showing the average change rate of the air quality based on the given length of time slot.
We call $\sigma_{d}^2$ as the temporal deviation variance.
Therefore the distribution of $X_{k,t+\tau}=X_{k,t}+X_d^{t\rightarrow t+1}+\cdots +X_d^{t+\tau-1 \rightarrow t+\tau}$ is given by
\begin{equation}\label{eqn_InferTimeError2}
\left\{\!\!\!\!
\begin{array}{ll}
 \!\!\!\!& X_{k,t+\tau} \sim\mathcal{N} (\mu_{k,t},\sigma_{k,t+\tau}^2), \\
 \!\!\!\!& \sigma_{k,t+\tau}^2 = \sigma_{k,t}^2+\tau\sigma_{d}^2,
\end{array}
\quad \forall t \ge 0, \, \forall k \in \mathcal{K},   \phi_{k,t}=1,
\right.
\end{equation}
which implies that the more time span it is, the less accurate the inference will be.

\textbf{Spacial inference by single source:} Based on $X_{k,t}$, no matter it is a directly measured value or the result of a temporal inference, we are able to infer the value at another location at the same time slot, $X_{k^\prime\!,t}$, as shown in Fig.~\ref{fig_SystemModel}.
To achieve this, we exploit the relevance among different locations from historical data and find that the deviations among different locations can also be modeled as additive.
Specifically, the additive random deviation from location $k$ to location $k^\prime$ is denoted as $X_d^{k\rightarrow k^\prime}$, obeying the following distribution:
\begin{equation}\label{eqn_InferSpace1Error1}
\left\{\!\!\!\!
\begin{array}{ll}
 \!\!\!\!& X_d^{k\rightarrow k^\prime}\sim\mathcal{N} (\mu_{k,k^\prime\!,t},\sigma_{k,k^\prime\!,t}^2), \\
 \!\!\!\!& \mu_{k,k^\prime\!,t} \approx \mu_t \times \mu_{k,k^\prime}, \\
  \!\!\!\!& \sigma_{k,k^\prime\!,t}^2 \approx \mu_t^2 \times \sigma_{k,k^\prime}^2,
\end{array}
\quad \forall t \ge 0, \forall k,k^\prime \in \mathcal{K}
\right.
\end{equation}
where $\mu_t$ is the average value at time $t$, $\mu_{k,k^\prime}$ is the constant describing the normalized average deviation from location $k$ to $k^\prime$, and $\sigma_{k,k^\prime}$ is the constant describing the normalized increased variance when using $X_{k,t}$ to infer $X_{k^\prime\!,t}$.
Also note that $\sigma_{k,k}^2\!=\!0$, $\forall k \in \mathcal{K}$.
Now we have the distribution of the inferred $X_{k^\prime\!,t}=X_{k,t}+X_d^{k\rightarrow k^\prime}$ as:
\begin{equation}\label{eqn_InferSpace1Error2}
\left\{\!\!\!\!
\begin{array}{ll}
 \!\!\!\!& X_{k^\prime\!,t}\sim\mathcal{N} (\mu_{k^\prime\!,t},\sigma_{k^\prime\!,t}^2), \\
 \!\!\!\!& \mu_{k^\prime\!,t} = \mu_{k,t} + \mu_{k,k^\prime\!,t}, \\
  \!\!\!\!& \sigma_{k^\prime\!,t}^2 = \sigma_{k,t}^2 + \sigma_{k,k^\prime\!,t}^2,
\end{array}
\quad \forall t \ge 0, \forall k,k^\prime \in \mathcal{K}.
\right.
\end{equation}
Note that as $\mu_t$ in (\ref{eqn_InferSpace1Error1}) gets larger (indicating worse air quality), the additional inference variance $\sigma_{k^\prime\!,t}^2$ in (\ref{eqn_InferSpace1Error2}) gets larger.

\textbf{Spacial inference by multiple sources:} We can further utilize $M$ values from multiple locations to infer an unknown value at a different location at the same time slot $t$.
The utilized values can either be the directly measured values or the inferred values through earlier measurement based on~(\ref{eqn_InferTimeError2}).
For each of these value, we use~(\ref{eqn_InferSpace1Error2}) to perform a single-source inference for the target location.
The $m^{th}$ inference result for the target location $k$ is denoted as $X_{k,t,m} \!\sim\! \mathcal{N} (\mu_{k,t,m},\sigma_{k,t,m}^2)$, where $1 \le m \le M$.
Then we can multiple all the probability density functions (PDF) of these inference results together to get the PDF of the target location.
For simplicity, we assume the distributions of different $X_{k,t,m}$ are independent (since they can be traced back to different sensors).
Therefore the final inference $X_{k,t}$ also has a Gaussian distribution, given as:
\begin{equation}\label{eqn_InferSpace2Error1}
\left\{\!\!\!\!
\begin{array}{ll}
 & X_{k,t} \sim \mathcal{N}(\mu_{k,t}, \sigma_{k,t}^2), \\
 & \mu_{k,t} = \dfrac{\sum\nolimits_{m=1}^M \mu_{k,t,m}/\sigma_{k,t,m}^2}{\sum\nolimits_{m=1}^M 1/ \sigma_{k,t,m}^2}, \\
 & \sigma_{k,t}^2 = \dfrac{1}{\sum\nolimits_{m=1}^M 1/\sigma_{k,t,m}^2},
\end{array}
\quad \forall t \ge 0, \forall k^\prime \in \mathcal{K}.
\right.
\end{equation}
where the inference result has a weighted mean based on the mean of these $M$ random variables and has a smaller variance compared with each one of these random variables.

\textbf{Rule of inference:}
For a measured value $X_{k,t}$ with $\phi_{k,t}\!=\!1$, no inference is performed.
For an unmeasured value $X_{k,t}$ with $\phi_{k,t}=0$, we consider a three-step inference.
The first step is to execute up to $L$ times of temporal inferences for all the selected locations based on their previous measured values, in which way we have $L$ intermediate results for the current time, according to Eqn.~(\ref{eqn_InferTimeError2}).
The second step is to utilize these intermediate results to perform $L$ times of ``single source" spatial inference for the target location, according to Eqn.~(\ref{eqn_InferSpace1Error2}).
And the final step is to combine these inferences to form a ``multi-source" spatial inference, according to Eqn.~(\ref{eqn_InferSpace2Error1}).
Fig.~\ref{fig_SystemModel} provides a simple illustration of the above inference steps.

\subsection{Environment Model}\label{sec_ModelEnvironment}

In the last subsection, we have mentioned $\mu_t$ as the average result of the $t^{th}$ time slot.
This value can be seen as the air quality for the whole area in a coarse-grained perspective.
Without the loss of accuracy, we consider this value is the same as the true average air quality for the whole area.
And we aim to establish a statistic model for the change of $\mu_t$.

From our collected data, we find that there is an approximately fixed statistic pattern of $\mu_t$.
Specifically, we can calculate how often does a certain level of polluted weather occurs, given by
\begin{equation}\label{eqn_AQIProb}
P[\mu_t=y],  \quad t \in [0,T], \quad  y\in\mathcal{Y},
\end{equation}
where $\mathcal{Y}$ is the value space of the possible air quality.
The values of air quality (such as PM2.5) are usually in the form of integer, thus we consider $\mathcal{Y}$ as a finite discrete value space.
In addition, for a fixed length of time slot (such as $10$ minutes), the probability of air quality transition between adjacent time slots can also be calculated, given by
\begin{equation}\label{eqn_AQIProbTrans}
P[\mu_t\!=\!y \big| \mu_{t-1}\!=\!y^\prime], \quad t \in [1,T], \quad y, y^\prime\in\mathcal{Y},
\end{equation}
where the current coarse-grained air quality $\mu_t$ has a relation with $\mu_{t-1}$.

It is assumed that $\mu_t$ can be roughly known when it comes to the $t^{th}$ time slot.
The corresponding approaches could be neural networks~\cite{bib_Prediction}, or checking the official weather report (which is not our focus in this paper).
We focus on how to increase the fine-grained air quality map by power control and location selection, as presenting in the next subsection.

\subsection{Problem of Power Control and Location Selection}

The limited capacity of each sensing device confines the number of sensing data it can collect.
For simplicity, we assume that the sensing devices have the same battery capacity and each one of them can only perform $E$ times of sensing tasks (including data sensing and an immediate data uploading) before its battery dies, where $E < T$.
Therefore, we have $\sum_{t=0}^{T}\phi_{k,t} \le E, \, \forall k \!\in\! \mathcal{K}_L$, showing the energy budget of the devices.
In addition, we expect that each device should not be silent for too long.
The maximum number of consecutive time slots that a device can keep asleep is $\Delta T$, which provides $\sum_{t}^{t+\Delta T+1} \phi_{k,t} \ge 1, \, \forall k \!\in\! \mathcal{K}_L, \, \forall t \!\in\![0,T\!-\!\Delta T\!-\!1]$.
We should guarantee that $\Delta T \cdot E >T$ to avoid contradiction.

Since the server needs to provide a \emph{real-time} distribution of the air quality, the incomplete data at the unmeasured locations should be inferred by the collected data according to the spatial and the temporal inference mentioned in Section~\ref{sec_ModelError}.
For a given time slot, when the current air quality map is established with the help of inference, we can investigate the accuracy of this map.
For $X_{k,t}$, we define its \emph{joint error}, $J_{k,t}$, as the indicator to quantitatively show reliability of the data, which is given as below:
\begin{equation}\label{eqn_Joint}
J_{k,t} = {\sqrt{\sigma_{k,t}^2+(\mu_{k,t}-\mu_{t})^2}},
\end{equation}
which jointly considers the variance of the value and the deviation from the current average value.
Specifically,  a larger variance or a larger deviation could increase the joint error of the data, i.e., $X_{k,t}$ is less reliable as $J_{k,t}$ gets larger.
Note that if $X_{k,t}$ is a measured value, then $\sigma_{k,t}^2=\mu_t\sigma_0^2$ and $\mu_{t,k}\approx\mu_t$.
We consider $\mu_{t,k}\!-\!\mu_t=0$ in this case for simplicity.
Otherwise, $\sigma_{k,t}^2$ and $\mu_{t,k}$ should be calculated according to Eqn.~(\ref{eqn_InferTimeError1})$\sim$(\ref{eqn_InferSpace2Error1}) based on the inference rule.
Either way, the joint error of each value at the current time slot can be calculated if we have determined the subset of sensing devices being turned on.

At the $t^{th}$ time slot, the average joint error of the current generated real-time air quality map is given by $\sum_{k\in \mathcal{K}} J_{k,t}/K$.
And for the whole period including $T$ time slots, the average joint error is calculated as
\begin{equation}
\bar{J}=T^{-1}K^{-1} \sum\limits_{t=1}^{T} \sum\limits_{k\in \mathcal{K}} J_{k,t},
\end{equation}
where we assume all the sensors should perform a sensing at $t=0$ for a good initialization and the situation at $t=0$ is not counted.
The objective function of minimizing the average joint error of the real-time air quality map is
\begin{align}\label{eqn_Objective}
\min\limits_{\mathcal{K}_L} & \min\limits_{\{\phi_{k,t}\}}  \quad \bar{J}, \\
s.t. &  \sum_{t=0}^{T}\phi_{k,t} \le E, \quad \forall k \in \mathcal{K}_L, \label{eqn_Constraint1}\\
&  \!\!\!\!\sum_{t}^{t+\Delta T+1} \!\!\!\!\phi_{k,t} \ge 1, \quad \forall k \in \mathcal{K}_L,  \forall t \in [0, T\!-\!\Delta T \!-\! 1], \label{eqn_Constraint2}\\
&  \phi_{k,t}=0,1 \quad \forall k \in \mathcal{K}_L,  \forall t \in [0, T], \\
&  \phi_{k,t}=0, \quad \forall k \in \mathcal{K}_U,  \forall t \in [0, T], \label{eqn_Constraint4}\\
& |\mathcal{K}_L|\le L, \quad |\mathcal{K}_L| \in \mathcal{K}, \label{eqn_Constraint5}\\
&  \mathcal{K}_L\bigcup\mathcal{K}_U=\mathcal{K}, \quad \mathcal{K}_L\cap \mathcal{K}_U = \varnothing, \label{eqn_Constraint6}
\end{align}
where Eqn.~(\ref{eqn_Constraint1})$\sim$(\ref{eqn_Constraint4}) show the constraints of power control and Eqn.~(\ref{eqn_Constraint5})$\sim$(\ref{eqn_Constraint6}) show the constraints of location selection.

\section{Theoretical Analysis}\label{sec_PowerControl}

In this section, we first take a deeper look into the three-step inference rule and obtain some basic properties of the joint inference in Section~\ref{sec_Theoretical_Sub1}.
Then we study the influence of the system parameters on the system performance in Section~\ref{sec_Theoretical_Sub2}.
Finally, we discuss some intuitions for the optimization problem in Section~\ref{sec_Theoretical_Sub3}, which leads to the solutions in Section~\ref{sec_PowerControl} and Section~\ref{sec_LocationSelection}.

\subsection{The Mean and The Variance of The Joint Inference}\label{sec_Theoretical_Sub1}

\begin{figure}[!thp]
\centering
\includegraphics[width=3.1in]{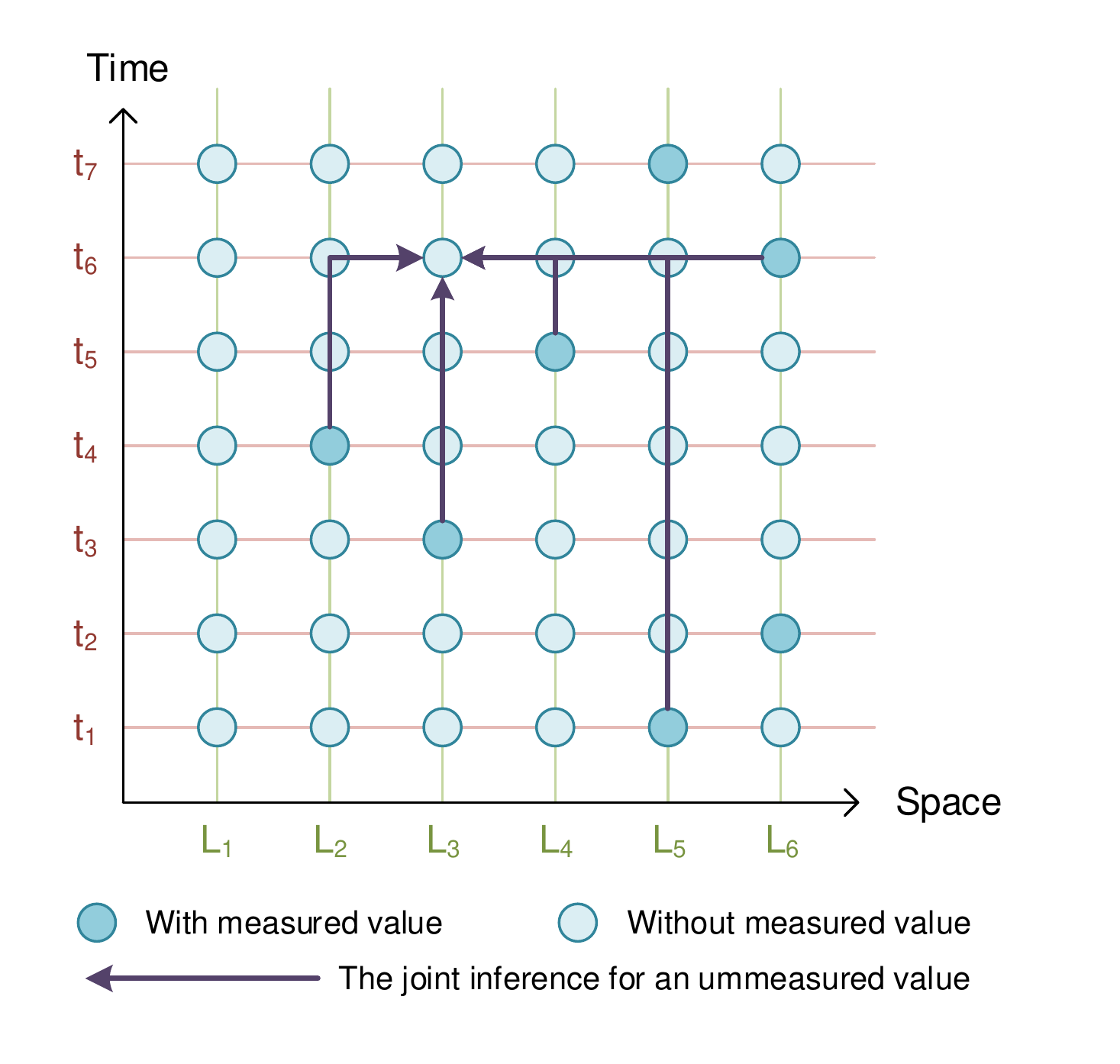}
\vspace{-3mm}
\caption{The joint inference for an unmeasured value in the spatial-temporal graph. The the given example, there are six locations and only five sensing device. Each unmeasured value is inferred by the most recent measured values from these five sensing devices.}\label{fig_InferenceRule}
\vspace{-3mm}
\end{figure}

From the three-step inference rule introduced in Section~\ref{sec_ModelError}, we know that each unmeasured value is inferred by $L$ values, which are the most current data that collected by the each one of the sensing devices.
We provide Fig.~\ref{fig_InferenceRule} as an example to illustrate such procedure.
The final inference is a multi-source spatial inference based on $L$ single-source spatial inferences.
And each single-source spatial inference is based on a temporal inference if this location has no current measured value.

Now we focus on the inference for a certain location $k_0$ at a certain time slot $t_0$.
We denote the time span that the $k^{th}$ device has not sense any data until $t_0$ as $\tau_k \in [0,\Delta T]$.
Therefore the $k^{th}$ intermediate inference result after the temporal and the single-source temporal inference for the target location $k_0$ is given by
\begin{equation}\label{eqn_Infer1}
\!\!X_{k_0,t_0,k} \!\sim\! \mathcal{N}(\mu_{t_0\!-\!\tau_k}+\mu_t\mu_{k,k_0}, \mu_{t_0\!-\!\tau_k}^2\sigma_0^2\!+\!\tau_k \sigma_d^2\!+\!\mu_t^2\sigma_{k,k_0}^2), \!\!\!
\end{equation}
where $\mu_{t_0\!-\!\tau_k}^2\sigma_0^2$ is the measurement variance, $\tau_k \sigma_d^2$ is the additional variance of temporal inference, and $\mu_t^2\sigma_{k,k_0}^2$ is the variance of spatial inference based on the relation of $k$ and $k_0$.
Since $\sigma_{k,k_0}^2\!=\!0$ if the variable $k\!=\!k_0$, the above expression is compatible for all situations, such as $(t_3,L_3)\!\!\rightarrow\!\!(t_6,L_3)$ in Fig.~\ref{fig_InferenceRule}.

To combine these $L$ results using a multi-source spatial inference, we use Eqn.~(\ref{eqn_InferSpace2Error1}) to calculate the mean value $\mu_{k_0,t_0}$ and the variance $\sigma_{k_0,t_0}^2$ of the final result.
For the convince of reading, we rewrite the expression of $\mu_{k_0,t_0}$ and $\sigma_{k_0,t_0}^2$ as below:
\begin{eqnarray}
\mu_{k_0,t_0} = &\!\!  \dfrac{\sum\nolimits_{k\in\mathcal{K}_L} \mu_{(k)}/\sigma_{(k)}^2}{\sum\nolimits_{k\in\mathcal{K}_L} 1/ \sigma_{(k)}^2}, \label{eqn_Infer2}\\
\sigma_{k_0,t_0}^2 = &\!\!  \dfrac{1}{\sum\nolimits_{k\in\mathcal{K}_L} 1/\sigma_{(k)}^2}, \label{eqn_Infer3}
\end{eqnarray}
where $\mu_{(k)}$ and $\sigma^2_{(k)}$ are short for the mean value and the variance of $X_{k_0,t_0,k}$, respectively, to facilitate reading in the rest of this section.

\begin{remark}
From Eqn.~(\ref{eqn_Infer2}), we can see that $\mu_{k_0,t_0} $ is the weighted sum of $\{\mu_{(k)}|k\in \mathcal{K}_L\}$.
The corresponding weight for the $k^{th}$ component is $(\sigma_{(m)}^2)^{-1}$, meaning that a more accurate single-source spatial inference affects more on the final result of the multi-source spatial inference.
In addition, we have $\min{\{\mu_{(k)}\}} < \mu_{k_0,t_0} < \max{\{\mu_{(k)}\}}$, since
\vspace{-2mm}
\begin{equation}
\mu_{min}\!=\!\dfrac{\sum\limits_{k\in\mathcal{K}_L} \!\!\dfrac{\mu_{min}}{\sigma_{(k)}^2}}{\sum\limits_{k\in\mathcal{K}_L} \!\!\dfrac{1}{ \sigma_{(k)}^2}}  \!\!<  \!\!\dfrac{\sum\limits_{k\in\mathcal{K}_L} \!\!\dfrac{\mu_{(k)}}{\sigma_{(k)}^2}}{\sum\limits_{k\in\mathcal{K}_L} \!\!\dfrac{1}{ \sigma_{(k)}^2}}  \!\!<  \!\!\dfrac{\sum\limits_{k\in\mathcal{K}_L} \!\!\dfrac{\mu_{max}}{\sigma_{(k)}^2}}{\sum\limits_{k\in\mathcal{K}_L} \!\!\dfrac{1}{ \sigma_{(k)}^2}} \!=\! \mu_{max},
\end{equation}
where $\mu_{min}=\min{\{\mu_{(k)}\}}$ and $\mu_{max}=\max{\{\mu_{(k)}\}}$.
\end{remark}

\begin{remark}
From Eqn.~(\ref{eqn_Infer3}), we can guarantee that $\sigma_{k_0,t_0}^2<\min{\{\sigma_{(k)}^2\}}$, since the following condition holds:
\vspace{-2mm}
\begin{equation}
\sigma_{(i)}^2 - \dfrac{1}{\sum\limits_{k\in\mathcal{K}_L} \dfrac{1}{\sigma_{(k)}^2}} =  \dfrac{\sum\limits_{k\in\mathcal{K}_L} \!\!\dfrac{\sigma_{(i)}^2}{\sigma_{(k)}^2} - \dfrac{\sigma_{(i)}^2}{\sigma_{(i)}^2}}{\sum\limits_{k=1}^L \dfrac{1}{\sigma_{(k)}^2}} >0, \,\,\,\, \forall i \in\mathcal{K}_L.
\end{equation}
\end{remark}

\begin{remark}
The final inference variance $\sigma_{k_0,t_0}^2$ is more sensitive to the minimal value of $\{\sigma_{(k)}^2\}$, since we have the following partial derivative:
\vspace{-2mm}
\begin{equation}
{\partial\bigg( \sum\limits_{k\in\mathcal{K}_L} \dfrac{1}{\sigma_{(k)}^2} \bigg)^{-1}}  \!\!\bigg/  {\partial(\sigma_{(i)}^2)} = \bigg( \sum\limits_{k\in\mathcal{K}_L} \dfrac{1}{\sigma_{(k)}^2} \bigg)^{-2} \!\!\!\!\!\!\cdot\! \big(\sigma_{(i)}^{2}\big)^{-2},
\end{equation}
which means that the same amount of decrease of a smaller $\sigma^2_{(i)}$ will lead to a larger reduce of the final variance.
\end{remark}

\subsection{Influence of The System Parameters on The Joint Error}\label{sec_Theoretical_Sub2}

From the expression of the joint error $J_{k,t}$ in Eqn.~(\ref{eqn_Joint}), we can see that both the variance of the inference result $\sigma_{k,t}^2$ and the deviation from the coarse-grained air quality $|\mu_{k,t}-\mu_t|$ contributes to $J_{k,t}$.
The increase of $\sigma_{k,t}^2$ and $|\mu_{k,t}-\mu_t|$ could decrease the inference accuracy and lower the confidence level of the established real-time air quality map.

From Eqn.~(\ref{eqn_Infer1}) and (\ref{eqn_Infer3}), we can see that the variance of the joint inference depends on the current air quality $\mu_t$, the time span $\tau_k$ since the most recent sensing, and the air quality $\mu_{t_0\!-\!{\tau_k}}$ when performing the most recent sensing.
This means that the temporal inference from a data long time ago (especially when the value was high back then) is questionable, and the spatial inference on a bad weather condition (high values of air quality) is also inaccurate.

From Eqn.~(\ref{eqn_Infer2}) and (\ref{eqn_Infer3}), we can see that the mean of the joint inference is the weighted mean of the corresponding values $\{\mu_{(k)}\}$ from all the $L$ sensing locations.
Since $\mu_{(k)}$ is actually the air quality of the $(t\!-\!\tau_k)^{th}$ time slot, its difference with the current value $\mu_t$ could be large if the air quality changes rapidly in the recent time slots.
From our statistical data mentioned in Section~\ref{sec_ModelEnvironment}, the air quality transition in adjacent time slots presents a greater probability for the similar air quality values, i.e., $P(\mu_t\!=\!y\big|\mu_{t-1}\!=\!y^\prime)$ is larger if $|y-y^\prime|$ is small.
This means that the air quality values in recent time slots is more reliable compared with the values from more previous time slots.
Thus $|\mu_{k,t}-\mu_t|$ is expected to be smaller if the sensing devices can turn on more frequently.

\begin{figure}[!thp]
\centering
\includegraphics[width=3.1in]{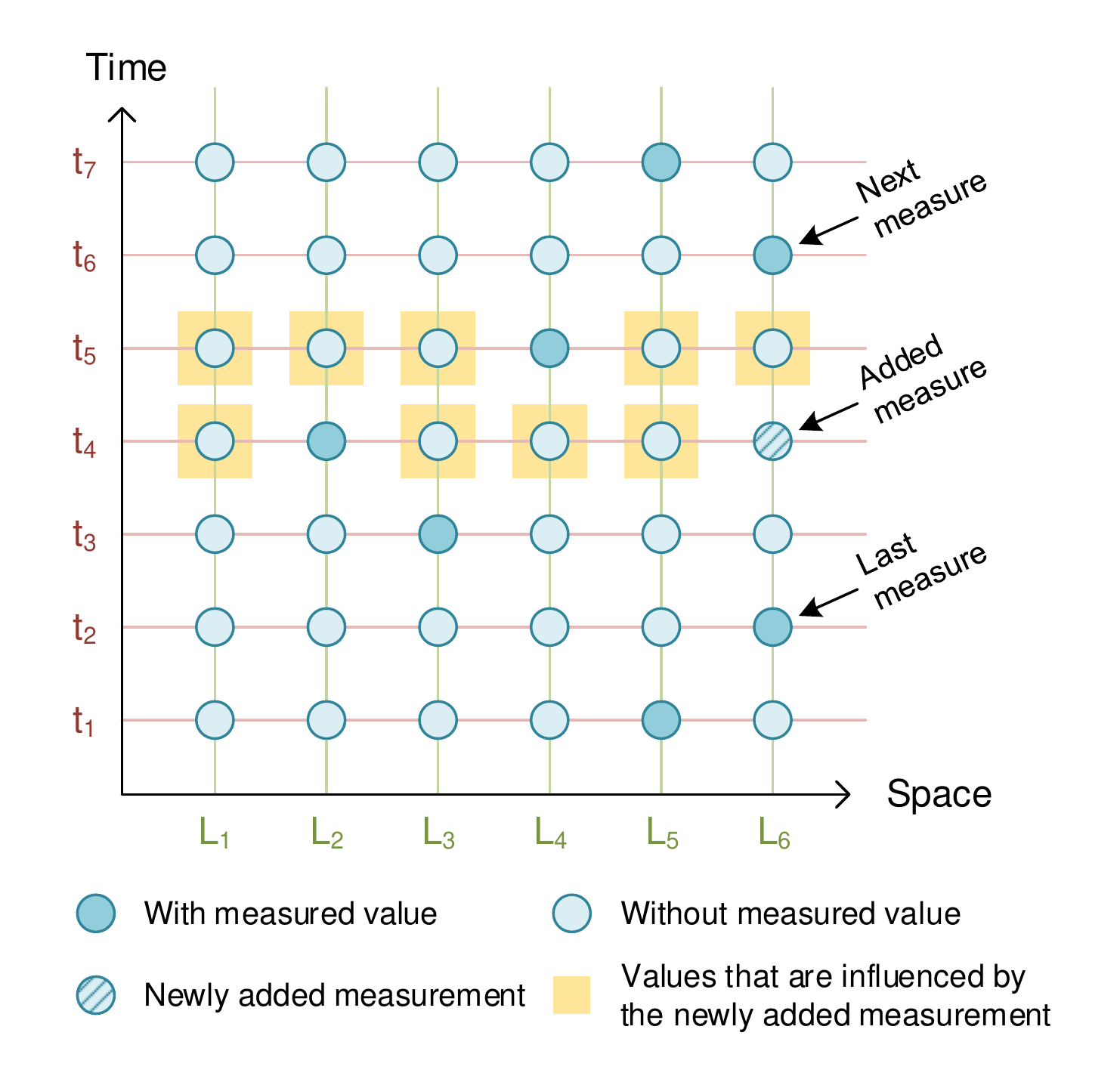}
\vspace{-3mm}
\caption{The influence of adding a new sensing point in the spatial-temporal graph. As we change a node from ``unmeasured value" to ``measured value", the joint error of some of the nodes are influenced.}\label{fig_Adding}
\vspace{-3mm}
\end{figure}

\begin{lemma}
Adding a measured value in the existing spatial-temporal graph of the air quality sensing system can averagely decrease the joint error.
\end{lemma}

\begin{IEEEproof}
We assume that the added measurement is at the $k^{th}$ location at the $t_2^{th}$ time slot, given by $X_{k,t_2}$.
And we denote the nearest measurement of location $k$ is at the $t_1^{th}$ and the $t_3^{th}$ time slots, with $t_1<t_2<t_3$.
As illustrated in Fig.~\ref{fig_Adding}, the influenced values are within $[t_2, t_3\!-\!1]$, where the earlier unmeasured values are inferred based on $X_{k,t_1}$ and the later unmeasured values are inferred based on $X_{k,t_3}$.
For each of these influence unmeasured values, $X_{k,t_2}$ provides a lower variance in the single-source spatial inference compared with the original value $X_{k,t_1}$ according to Eqn.~(\ref{eqn_Infer1}).
This is because the value of $\tau_k$ is smaller and the probability distribution of $\mu_{t_2}$ is the same as $\mu_{t_1}$ in the long term average observations.
In addition, $|\mu_t-\mu_{t_2}|$ also has a smaller expectation than $|\mu_t\!-\!\mu_{t_1}|$ for all $t\in[t_2,t_3\!-\!1]$ since the aforementioned property of statistical air quality transition.
\end{IEEEproof}

Note that the conclusion of Lemma 1 shows the average outcome of the situations.
Based on Lemma 1, we can directly obtain the following propositions:

\begin{proposition}\label{pro_Power}
Given a fixed time period $T$, a fixed number of sensing devices $|\mathcal{K}_L|$, two different settings of energy budget $E_1<E_2$, the corresponding average joint errors comply to $\bar{J}_1\ge \bar{J}_2$ in the optimal power control strategy.
\end{proposition}

\begin{IEEEproof}
We assume that the best power control strategy of $E_1$ is $\{\phi_{k,t}\}$, where $\sum\nolimits_{t}\phi_{k,t}<E_1$, $\forall k \in \mathcal{K}$.
As we raise the energy budget from $E_1$ to $E_2$, more values of $\phi_{k,t}$ can be changed from $0$ to $1$.
Based on Lemma 1, adding a new measured value can averagely reduce the average joint error.
Even in the worst case where no newly added measurement increases inference accuracy due to extreme weather condition, we can keep $\phi_{k,t}$ as it is and do not deteriorate the original result.
\end{IEEEproof}

\begin{proposition}\label{pro_DeviceNumber}
Given a fixed time period $T$, a fixed energy budget $E$, two different settings of the number of available sensing devices $L_1<L_2$, the corresponding average joint errors comply to $\bar{J}_1 \ge \bar{J}_2$ in the optimal power control and location selection strategy.
\end{proposition}

\begin{IEEEproof}
We assume the optimal power control and location selection for $L_1$ devices are $\{\phi_{k,t}\}$ and $\mathcal{K}_L$, respectively.
Assume that we add one more device at location $k^\prime$, then its collected data can be used to infer the values at the unselected locations for $t\in[0,T]$, and the values at its own location only for $\{t \,|\, \phi_{k^\prime,t}=0\}$.
From Eqn.~(\ref{eqn_Infer3}) we know that the variance of the inference decreases since an additional value participates in the multi-source inference.
The remaining problem is to figure out how $|\mu_{k,t}-\mu_t|$ of each inferred value changes.
A basic idea is to let the newly added device to copy the power scheduling of one of the existing device.
According to Remark~1, this is equivalent to the action of adding the weight of the copied device when calculating Eqn.~(\ref{eqn_Infer2}).
It is expected that some of the $\{|\mu_{k,t}-\mu_t|\}$ will increase and some will decrease.
Find the best existing device to copy its power scheduling can averagely achieve positive effect, which will generally reduce $\bar{J}$.
Even in the worst case where the newly added device results in a worse $\bar{J}$ due to some extreme settings, we can eliminate the newly added device and keep the original location selection plan, resulting a same $\bar{J}$.
\end{IEEEproof}

\subsection{Discussions on The Formulated Optimization Problem}\label{sec_Theoretical_Sub3}

For the location selection, intuitively, the devices need to be deployed in those less correlated locations (with high values of $\sigma^2_{k,k^\prime}$ between each other), acquiring ``more diversified" data to help re-establish the fine-grained air quality map.

For the power control, the turning-on frequency of the sensing devices should be properly adjusted.
A low frequency sensing plan could reduce the accuracy of the real-time air quality map, and a too frequent sensing plan may lead to the the depletion of their batteries long before the last hour $T$.

It should be noted that, both the measurement and inference error depends on the average air quality ($\mu_t$).
This means that we need to know the air quality in advance to make the perfect strategy, which is not a acceptable assumption.
We aim to create a more generalized power control strategy which can dynamically deal with the encountered weather condition as long as the statistics of the air quality ($P[\mu_t=y]$ and $P[\mu_t=y|\mu_{t-1}=y^\prime]$) is fixed.
Therefore, we only assume the current and the previous air quality ($\mu_t$, $t\in[0,t_{now}]$) is known as the system is establishing the air quality map at $t_{now}$.

In fact, the joint optimization of power control and location selection is highly intractable even with the help of the statistics of historical data.
Therefore, in the following part of this paper, we separate problem into the power control problem and the location selection problem.
Specifically, we first study the problem of power control in a stochastic environment based on a fixed location selection in Section~\ref{sec_PowerControl}.
Next, in Section~\ref{sec_LocationSelection}, we study the problem of location selection based on a fixed power control strategy in a given environment.
By combining the solutions for these two individual problems together, its is expected that a satisfactory outcome can be acquired.

\section{Power Control Strategy}\label{sec_PowerControl}

In this section, we provide the power control strategy with a fixed location selection $\mathcal{K}_L$.
With the knowledge of the environment statistics (as $P[\mu_t\!=\!x]$ and $P[\mu_t\!=\!x \big| \mu_{t-1}\!=\!x^\prime]$), we aim to provide a best power control strategy that is able to deal with the unknown environment having the same statistics.
In our context, the power control strategy is learnt by means of reinforcement learning.

However, before formally studying the problem of multiple devices, we first take a look at a simpler situation where only one device is included.
Analyzing and solving this simpler problem can help us deal with the case of multiple devices.
Specifically, the problem of power control for a single device can be transformed into a Markov Decision Process (MDP), and solved by a dynamic programming algorithm optimally, as provided in Section~\ref{sec_PowerControlSingle}.
Since the complexity of the optimal dynamic programming algorithm increases exponentially with the number of devices, we provide a deep Q-learning solution with approximated value functions for the problem of multiple devices in Section~\ref{sec_PowerControlMultiple}.

\subsection{Power Control for Single Device}\label{sec_PowerControlSingle}

In this subsection, we assume that the number of available device is one, i.e., $L\!=\!1$.
This means that all the efforts of the power control is concentrated on this single device.
In the following, we establish a MDP model with discrete and finite state space, which describes the state transition during the power control procedure.

A MDP consists of five components, namely, the set of states $\mathcal{S}$, the set of available actions $\mathcal{A}$, the state transition probability matrix $\mathcal{P}$, the reward function $\mathcal{R}$, and the discount factor $\gamma$.
To be specific, the states in $\mathcal{S}$ should obey the Markov property, where each next state only depends on the current state and the adopted action.
Assume that the current state is $s$, one can choose an action $a$ from the action set $\mathcal{A}$ to make the system change.
There could be multiple consequent states $\{s^\prime\}\in\mathcal{S}$ after performing $a$ on $s$, and the corresponding transition probability is given by $\mathcal{P}_{ss^\prime}^a \!=\! \big[S_{i+1}\!=\!s^\prime \big| S_i\!=\!s, A_i \!=\! a\big]$, where $S_i$ and $A_i$ represents the $i^{th}$ state and the $i^{th}$ action in the whole history.
In addition, there is an reward $\mathcal{R}_s^a$ of performing $a$ on $s$, representing the immediate utility/gain.
The discount factor $\gamma \!\in\! [0,1] $ indicates the fading utility of the future rewards from the viewpoint the current state.

\begin{definition}
In the power control problem with a single sensing device, the $i^{th}$ system state in the whole state transition history is defined in the following form:
\begin{equation}
S_i=(S^t_i,S^p_i,S^d_i,S^r_i,S^e_i), \label{eqn_StateSingle}
\end{equation}
which has five components. The integer $S^t_i \!\in\! [0,T\!+\!1]$ represents the time of the system (``t" for ``time"). The integer $S^p_i \!\in\! [0,E]$ indicates the remaining power of the sensing device (``p" for ``power"). The integer $S^d_i \in [0,\Delta T]$ shows the number of time slots since the last time of measurement (``d" for ``delay"). The integer $S^r_i \!\in\! \mathcal{Y}$ records the average air quality value during the last time of measurement (``u" for ``record"). And the integer $S^e_i \!\in\! \mathcal{Y}$ shows the current average air quality $\mu_t$  (``e" for ``environment").
\end{definition}

\textbf{Initial state:} The initial state is given by $S_1\!=\!(1,E,1,\mu_0,\mu_1)$, which means that it is the $1^{th}$ time slot, and there are $E$ available chances of sensing.
Note that since we assume all the devices perform a sensing as soon as being deployed at the $0^{th}$ time slot (which is not counted in the energy budget), the time span since the last sensing is $S_1^d=1$, and the recorded air quality is $S_1^r\!=\!S_0^e\!=\!\mu_0$.

\begin{figure}[!thp]
\centering
\includegraphics[width=3.5in]{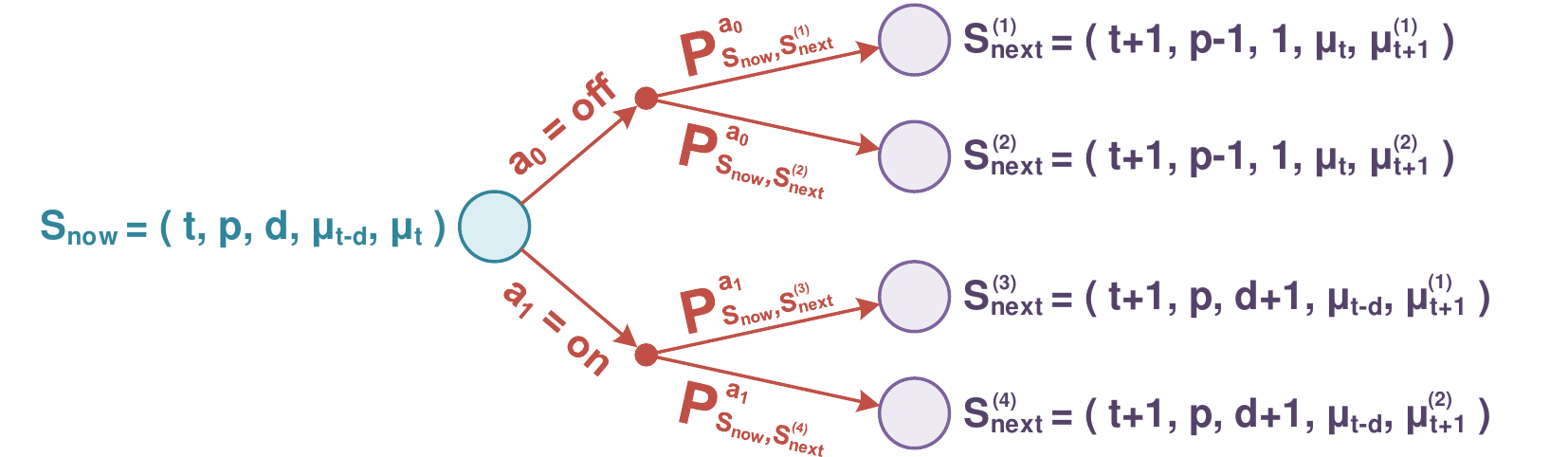}
\vspace{-3mm}
\caption{An illustration of the state transition. Each action may lead to multiple subsequent states since the environment change is random.}\label{fig_State}
\vspace{-2mm}
\end{figure}

\textbf{Action set:} From any given intermediate state, $S_i = (S^t_i,S^p_i,S^d_i,S^r_i,S^e_i)$, $\forall 1\!\le\! i \!\le\! T$, two actions can be performed.
Specifically, the action set is given by $\mathcal{A}=\{a_0,a_1\}$, where $a_0$ is to keep the sensing device asleep and $a_1$ is to turn on the sensing device.
If $S^p_i>0$ and $S^d_i=\Delta T$, then only $a_1$ is available since $\Delta T$ is the maximum time to keep a device asleep.
If $S^p_i=0$, then only $a_0$ is available since the power has been depleted\footnote{If $S^p_i=0$ and $S^d_i=\Delta T$, we only perform $a_0$ and force $S^d_{i+1}$ to be $\Delta T$ instead of $\Delta T\!+\!1$ to limit the number of states being represented. The occurrence of $S^p_i=0$ and $S^d_i=\Delta T$ represents the violation of the constraint Eqn.~(\ref{eqn_Constraint2}), indicating that state $S_i$ involves improper power control at previous steps, which will not be included in the optimal solution.}.
Note that although the subscript of each state, $i$, equals to $S_i^t$ in the single device problem, we distinguish them as two variables for the preparing for the multi-device problem.
Fig.~\ref{fig_State} shows an example of the actions performed on each state.

\textbf{Performing ``off" action:} If we perform $a_0$ on state $S_i$, it means that we execute no sensing task at time $S^t_i$.
The overall joint error $\sum_{k\in\mathcal{K}}  J_{k,{S^t_i}}$ at the $({S^t_i})^{th}$ time slot can be calculated according to Eqn.~(\ref{eqn_Infer1}) by setting $\mu_{t}\!=\!S^e_i$, $\tau\!=\!S^d_i$ and $\mu_{t\!-\!\tau}\!=\!S^r_i$.
We define the reward of taking action $a_0$ on state $S_i$ as the opposite value of $\sum_{k\in\mathcal{K}}  J_{k,{S^t_i}}$, written as
\begin{equation}
R_{S_i}^{a_0}= -\sum_{k\in\mathcal{K}}  J_{k,{S^t_i}} \Big|_{(S^e_i,S^d_i,S^r_i)}.
\end{equation}
The following state will be $S_{i+1}\!=\!(S^t_i\!+\!1,S^p_i,S^d_i\!+\!1,$ $S^r_i,S^e_{i\!+\!1})$.
Note that the first four components are determined, and the last component $S_{i\!+\!1}^e$ is generated randomly according to the air quality transition probability.
We denote the probability of $S_i$ changing to state $S_{i\!+\!1}$ by taking action $a_0$ as $\mathcal{P}_{S_i,S_{i\!+\!1}}^{a_0} \!\!= P[\mu_{t\!+\!1}\!=\!S_{i\!+\!1}^e\big|\mu_t\!=\!S_{i}^e]$.

\textbf{Performing ``on" action:} If we perform $a_1$ on state $S_i$, it means that we perform a sensing task at time $S^t_i$.
The overall joint error $\sum_{k\in\mathcal{K}}  J_{k,{S^t_i}}$ at the $({S^t_i})^{th}$ time slot can be calculated according to Eqn.~(\ref{eqn_Infer1}) by setting $\mu_{t}\!=\!S^e_i$, $\tau\!=\!0$ and $\mu_{t\!-\!\tau}\!=\!S^e_i$.
The corresponding reward is
\begin{equation}
R_{S_i}^{a_1}= -\sum_{k\in\mathcal{K}}  J_{k,{S^t_i}} \Big|_{(S^e_i,0,S^e_i)}.
\end{equation}
And the following state will be $S_{i+1}\!=\!(S^t_i\!+\!1,S^p_i\!-\!1,1, S^e_i,$ $S^e_{i\!+\!1})$, where $S^d_{i+1}\!=\!1$ because it has been one time slot since the last time of sensing, $S^r_{i+1}\!=\!S^e_i$ indicates the recorded air quality when performing sensing, and $S_{i\!+\!1}^e$ also complies to the air quality transition probability based on $S_{i}^e$.
We denote the probability of $S_i$ changing into $S_{i\!+\!1}$ by taking action $a_1$ as $\mathcal{P}_{S_i,S_{i\!+\!1}}^{a_1} \!\!= P[\mu_{t\!+\!1}\!=\!S_{i\!+\!1}^e\big|\mu_t\!=\!S_{i}^e]$.

\textbf{Termination condition:} It can be seen that no matter we use action $a_0$ or $a_1$, the component $S_i^t$ increases by one at each time of the state transition.
When it comes to $S_i^t=T$, we need to make the last action and the subsequent state will be $(T\!+\!1,S^p_i,S^d_i,S^r_i,S^e_i)$, which shows the termination of the state transition.

\textbf{State-value function:} For each state, there is a value function representing the utility of this state, denoted by $V(S)$.
Specifically, the termination state $(T\!+\!1,S^p_i,S^d_i,S^r_i,S^e_i)$ has zero utility, given by $V(S_{i+1})=0$.
In each intermediate step, if $S_i\rightarrow S_{i\!+\!1}$ with reward $R_{S_i}^{a}$ (with $a=a_0$ or $a=a_1$), then we have
\begin{equation}\label{eqn_ValueStateFunction}
V(S_i) = R_{S_i}^a + \gamma V(S_{i+1}),
\end{equation}
where the discount factor $\gamma$ is set to $1$ in our calculation.
It can be seen that the value of $S_1$ is the sum of the rewards along the path of the experienced states, given by
\begin{equation}\label{eqn_vs1}
V(S_1) = \sum\limits_{i=1}^T R_{S_i}^a = -\sum\limits_{t=1}^T\sum\limits_{k\in\mathcal{K}} J_{k,t},
\end{equation}
where we can see that maximizing $V(S_1)$ is the same as minimizing the average joint error $\bar{J}$ as the objective function describes.

\textbf{Action strategy:} The problem of maximizing $V(S_1)$ is to find a best path in the state space, which has a size of $T\!\times\! E\!\times\! \Delta T \!\times\! |\mathcal{Y}|^2$.
Since the state transition is not fixed due to the random change of air quality, the problem can be interpreted as how to decide the action for each possible state, given by
\begin{equation}
\pi(s) \in \{a_0, a_1\}, \quad \forall s.
\end{equation}
As proved in\cite{bib_Reinforce}, there exists an optimal deterministic action strategy for MDP.
That is to say, the optimal action strategy $\pi(s)$ for any given state does not need to be a probatilistic one (e.g. with $1/3$ probability choosing $a_0$ and with $2/3$ probability choosing $a_1$).

\textbf{Dynamic programming algorithm:} The MDP of the single device problem is highly structured.
Each state with $S_i^t$ can only change to another state with $S_i^t\!+\!1$, indicating an unidirectional dependence of the states.
Since all the termination states with $S_i^t\!=\!T\!+\!1$ have zero value, we can iteratively use the values of the state with $S_{i\!+\!1}^t$ to calculate the values of the state with $S^t_i$.
Specifically, we have
\begin{align}
& V(S_i)  = \!\!\max\limits_{a=a_1,a_0} \bigg[ R_{S_i}^a \!\!+\!\! \sum_{S_{i\!+\!1}} \mathcal{P}_{\!\!S_i,S_{i\!+\!1}}^a V(S_{i+1}) \bigg], \label{eqn_Dynamic1}\\
& \pi(S_i) = arg\!\!\max\limits_{a=a_1,a_0} \bigg[ R_{S_i}^a \!\!+\!\! \sum_{S_{i\!+\!1}} \mathcal{P}_{\!\!S_i,S_{i\!+\!1}}^a V(S_{i+1}) \bigg],  \label{eqn_Dynamic2}
\end{align}
where we should calculate $V(S_{i\!+\!1})$ for all possible $S_{i\!+\!1}$ before calculating $V(S_{i})$.
Since each value of $V(S_i)$ considers all the possible subsequent states, $V(S_i)$ can be maximized and the corresponding $\pi(S_i)$ is the optimal choice for the state $S_i$.
At the end of the iteration procedure, we acquire the final optimal strategy $\{\pi(s)\}$ for all the possible states.
Therefore, we can use $\{\pi(s)\}$ to deal with the single-device power control problem in an online mode, where the actions can dynamically adapt to the randomly changed environment ($\mu_t$).

\begin{algorithm}[!thp]
\caption{Optimal single-device power control.}\label{alg_Single}
\KwIn{Measurement variance $\sigma_0^2$, temporal inference variance $\sigma_d^2$, spatial inference variance $\{\sigma_{k,k^{\prime}}^2\}$, and air quality transition matrix $P_{|\mathcal{Y}|\times|\mathcal{Y}|}$.}
\KwOut{Optimal state-action strategy $\{\pi(s)\}$.}
\Begin
{
Initialize $V(s)=0$ for $s=(T\!+\!1,p,u,r,e)$, $\forall p\in[0,E], \forall u \in[1,\Delta T], \forall r,e\in\mathcal{Y}$\;
\For{$t$ is from $T$ to $1$}
     {
     Calculate $V(s)$ and $\pi(s)$ with $s=(t,p,u,r,e)$, $\forall p\in[0,E], \forall u \in[1,\Delta T], \forall r,e\in\mathcal{Y}$ according to~(\ref{eqn_Dynamic1}) and (\ref{eqn_Dynamic2})\;
     }
}
\end{algorithm}

\textbf{Computation complexity:} The value of each state is calculated once.
And to calculate the value of each state, no more than $|\mathcal{Y}|$ subsequent states are being considered.
Therefore, the final computation complexity is $O(T\!\cdot\!E\!\cdot\!\Delta T\!\cdot\!|\mathcal{Y}|^3)$.
If the value space of the air quality $|\mathcal{Y}|$ can be approximated into multiple segments, the complexity can greatly reduce.
An overview of this solution is presented in Algorithm~\ref{alg_Single}.

\subsection{Power Control for Multiple Devices}\label{sec_PowerControlMultiple}

For the problem of $L$ devices, the intuition is to define the MDP states by extending the one in (\ref{eqn_StateSingle}).
Specifically, we have
\begin{equation}\label{eqn_Multiple}
S_i=(S^t_i,\vec{S^p_i},\vec{S^d_i},\vec{S^r_i},S^e_i),
\end{equation}
where $\vec{S^p_i}$, $\vec{S^d_i}$, and $\vec{S^r_i}$ are the row vectors with length $L$, representing for all the $L$ devices.
The possible action for each state is also a $L$-length vector, given by $\big(a(1),a(2),\cdots,a(L)\big)$, where $a(l)\!=\!a_0$ or $a_1$, $\forall l \!\in\![1,L]$.
It is easy to see that the number of states is $T\!E^L\!(\Delta T)^L\!|\mathcal{Y}|^{L+1}$, and the number of actions is $2^L$.
Therefore, the optimal dynamic programming algorithm is no longer suitable to solve the multi-device power control problem.

Since both the extremely large state space and value space pose challenge for solving the problem, we first aim to transfer the complexity of the value space to the complexity of the state space.
This is done by arranging the sensing devices to take actions in a predefined order.
In this way, there are only two possible actions ($a_0$ and $a_1$) for each state.
And the number of states will be multiply by $L$ after such arrangement.

\begin{definition}
In the power control problem with $L$ sensing devices, the $i^{th}$ system state in the whole state transition history is defined in the following form:
\begin{equation}
S_i=(S^t_i,\vec{S^p_i},\vec{S^d_i},\vec{S^r_i},S^e_i,S^l_i), \label{eqn_StateMultiple}
\end{equation}
which has six components. The integer $S^t_i \!\in\! [0,T\!+\!1]$ represents the time of the system. The $L$-length integer vector $\vec{S^p_i}$ indicates the remaining power of each sensing device, with $\vec{S^p_i}(l)\!\in\! [0,E]$, $\forall 1\!\le\! l \!\le\! L$. The $L$-length integer vector $\vec{S^d_i}$ shows the number of time slots since the last time of measurement for each device, with $\vec{S^d_i}(l)\!\in\! [0,\Delta T]$, $\forall 1\!\le\! l \!\le\! L$. The $L$-length integer vector $\vec{S^r_i}$ records the average air quality value during the last time of measurement for each device, $\vec{S^r_i}(l)\!\in\!\mathcal{Y}$, $\forall 1\!\le\! l \!\le\! L$. The integer $S^e_i \!\in\! \mathcal{Y}$ shows the current average air quality $\mu_t$. And the integer $S^l_i \!\in\! [1,L]$ implies who's turn it is to take the action at this state.
\end{definition}

\textbf{Initial state:} The initial state is given by $S_1=(1,\vec{S^p_1},\vec{S^d_1},\vec{S^r_1},\mu_1,1)$, where $\vec{S^p_1}(l)=E$, $\vec{S^d_1}(l)=1$, $\vec{S^r_1}(l)=\mu_0$, $\forall 1\le l \le L$.
Note the last component of $S_1$ is $1$, indicating that it is the turn of the $1^{st}$ device to take action.

\textbf{Alternation rule:} The first component $S^t_i$ and the last component $S^l_i$ of each state obey the following rule in the state transition process, regardless of the exact actions being performed.
If we have $S^l_i<L$, then $S^l_{i+1}=S^l_i+1$ and $S^t_{i+1}=S^t_i$, meaning that it is the turn of the next device to decide sensing or not in the same time slot.
Otherwise, $S^l_{i+1}=1$ and $S^t_{i+1}=S^t_i+1$, indicating that all the devices have done making decisions in the current time slot and the time moves on.

\textbf{Action set:} For each intermediate state $S_i=(S^t_i,\vec{S^p_i},\vec{S^d_i},\vec{S^r_i},S^e_i,S^l_i)$, two actions can be performed, given by $\mathcal{A}=\{a_0,a_1\}$.
If $S^p_i(S^l_i)>0$ and $S^d_i(S^l_i)=\Delta T$, meaning that the $(S^l_i)^{th}$ device has been asleep long enough and still have power to perform sensing, then only action $a_1$ can be executed.
If $S^p_i=0$, meaning that the $(S^l_i)^{th}$ device has no power, then only $a_0$ can be executed.
For other cases, both $a_0$ and $a_1$ can be chosen for the $(S^l_i)^{th}$ device.

\textbf{State transition for $S^l_i<L$:} Assume that the current state is $S_i=(S^t_i,\vec{S^p_i},\vec{S^d_i},\vec{S^r_i},S^e_i,S^l_i)$.
If $a_0$ is performed, then $S_{i\!+\!1}^l=S^l_i+1$, with other components the same as $S_i$.
If $a_1$ is performed, then $S^l_{i\!+\!1}=S^l_i+1$, $S_{i\!+\!1}^p(S_i^l)=S_{i}^p(S_i^l)-1$, $S_{i\!+\!1}^d(S_i^l)=0$, $S^r_{i\!+\!1}(S_i^l)=S^e_i$, with other components the same as $S_i$.

\textbf{State transition for $S^l_i=L$:} Assume that the current state is $S_i=(S^t_i,\vec{S^p_i},\vec{S^d_i},\vec{S^r_i},S^e_i,S^l_i)$.
If $a_0$ is performed, then $S_{i\!+\!1}^t = S_i^t+1$, $S_{i\!+\!1}^l = 1$, $\vec{S^d_{i+1}}=\vec{S^d_i}+1$, $S^e_{i\!+\!1}\!\sim\! P[\mu_t\big|\mu_{t-1}\!=\!S^e_i]$, with other components the same as $S_i$.
If $a_1$ is performed, then $S_{i\!+\!1}^t = S_i^t+1$, $S_{i\!+\!1}^l = 1$, $S_{i\!+\!1}^p(L)=S_{i}^p(L)-1$, $S^r_{i\!+\!1}(L)=S^e_i$, $S^d_{i\!+\!1}(l)=S^d_i(l)+1$ for $1\!\le\! l\!\le\! L-1$,  $S^d_{i\!+\!1}(L)=1$, $S^e_{t\!+\!1}\!\sim\! P[\mu_t\big|\mu_{t-1}\!=\!S^e_i]$, with other components the same as $S_i$.

\textbf{Reward:} Unlike the case for the single-device problem, there are $L$ states for each time slot  as we experience through states.
It is not reasonable to still use the sum of the joint error $\sum_{k\in\mathcal{K}}  J_{k,{S^t_i}}$ of the current state-action pair as the corresponding reward.
This is because only the state in which all the $L$ devices has taken their actions contributes to the final joint error.
Therefore, we define the rewards of the state transition within one time slot as their marginal gain of taking actions.
Specifically, for the state with $S^l_i=1$, the reward is defined as $-\sum_{k\in\mathcal{K}}  J_{k,t}(1)$, representing the current joint error after the $1^{th}$ device take its action.
For reading convenience, we do not put the complicated expression here.
For each $S^l_i>1$, the reward is defined as the marginal decrease of the joint error in this time slot, given by $\sum_{k\in\mathcal{K}}  J_{k,t}(S^l_i-1)-\sum_{k\in\mathcal{K}}  J_{k,t}(S^l_i)$.
When it comes to the last state of this time slot, the sum of these $L$ rewards equals to $-\sum_{k\in\mathcal{K}}  J_{k,t}(L)$, which corresponds to the value of the joint error after all the $L$ devices take their actions.
Furthermore, the total reward in the whole state transition procedure actually equals to the opposite value of the summation of the joint error for all time slots, just like Eqn.~(\ref{eqn_vs1}).

\begin{proposition}\label{pro_Equivalent}
The optimal solution for the MDP in (\ref{eqn_Multiple}) and the optimal solution for the MDP in (\ref{eqn_StateMultiple}) achieve the same performance (the same $\bar{J}$).
\end{proposition}

\begin{IEEEproof}
Both MDP problems can be optimally solved by the dynamic programming algorithm (regardless of their complexity).
In the optimal solution of MDP, $V(s)$ can be maximized for all possible $s$.
And for the initial state $S_1$, $V(S_1)$ equals to the sum of the rewards along the path (also equals to the opposite value of the overall joint error).
Therefore, the optimal solutions for these two MDPs have the same performance, given as $\bar{J}=-V(S_1)/T/K$.
\end{IEEEproof}

Proposition~\ref{pro_Equivalent} indicates that the method of arranging the sensing devices to take actions in a predefined order is feasible.
However, we should also cope with problem of the large number of states, since it would take too long to figure out $V(s)$ for each state.
One possible solution in the reinforcement learning to deal with such a situation is to use an approximation function to estimate the value of each state, or to estimate the value of each state-action pair~\cite{bib_AlgoRein}.
Specifically, the value of performing action $a$ on state $s$ is denoted as $Q(s,a)$.
The relation of $V(s)$ and $Q(s,a)$ is shown as follows:
\begin{align}
& V(s) = \max\limits_a \big[ Q(s,a) \big], \label{eqn_vs}\\
& Q(s,a) = R_s^a + \sum\limits_{s^\prime} \mathcal{P}_{ss^\prime}^a V(s^\prime). \label{eqn_qsa}
\end{align}
By substituting Eqn.~(\ref{eqn_vs}) into Eqn.~(\ref{eqn_qsa}) and using the mathematical expectation instead of probability matrix, we have
\begin{equation}\label{eqn_qrq}
Q(s,a) = \mathbb{E}_{s^\prime}\big[ R_s^a + \max_{a^\prime} Q(s^\prime, a^\prime) \big].
\end{equation}

\textbf{Approximation of Q(s,\,a):} We denote the approximated $Q(s,a)$ as $\tilde{Q}(s,a)$, which should be accurate enough to tell the difference of taking different actions at a certain state.
The approximation function could be a linear one, e.g., $\tilde{Q}(s,a)=\sum_i w_i \!\cdot\! f_i(s,a)$, where $f_i(s,a)$ is the $i^{th}$ feature of the state-action pair, and all the features constitute a feature vector $\vec{f}(s,a)$.
However, the linear form of the approximation may have bad fitting performance due to the complexity of the state-action space.
Therefore, we use a deep neural network instead, which uses features as input and calculates the corresponding approximated value $\tilde{Q}(s,a)$, also known as deep Q-learning.
Here we denote the neural network model as $\tilde{Q}(s,a) = \mathcal{NN}\big(\vec{f}(s,a)\big)$.

\textbf{Feature design:} Since these features are used to train the expression of $\tilde{Q}(s,a)$, a careful design is necessary.
We first define the power deficiency of the device, given by
\begin{equation}
PD(t,p)=1\big/\big(1+\exp^{T/E-(T-t)/p}\big),
\end{equation}
where $T\!-\!t$ and $p$ are the remaining time and the remaining power, respectively.
If $(T\!-\!t)/p$ is larger than $T/E$, then $0.5\!<\!PD\!<\!1$, indicating the power is not sufficient compared with the remaining time.
Otherwise $0\!<\!PR\!<\!0.5$, showing the power is relatively sufficient compared with the remaining time.
A higher PD indicates the device is less willing to turn on and may relay on the measurement of other devices.
In our implementation, the length of the feature vector is $5L+K+3$, where $L$ is the number of devices and $K$ is the number of all the locations.

\textbf{Feature specifics:} To facilitate explanation, we consider the feature vector as five segments.
The first segment has $L$ components as $0$-$1$ values, with the $(S^l_i)^{th}$ location set as $1$ and all other locations set as $0$.
This is to indicate the specific device that is responsible to make an action in this state.
The length of the second segment is $L$, which contains the power state vector $\vec{S^p_i}$, showing the remaining power of each device.
The third segment has $L\!+\!K\!+\!1$ components, they are set to be zeros if $a=a_0$, otherwise, they show the marginal utility of taking action $a_1$.
Specifically, in the third segment, we have $K$ components showing the decreases of the joint error of all these $K$ location if performing $a_1$, we have additional $L$ components showing the weighted (multiplied by $PD$) decrease of the joint error for these $L$ devices, and another constant component $1$ just to indicate the action of $a_1$.
The fourth segment has $2L\!+\!1$ components, they are set to be zeros if $a=a_1$, otherwise, they show the potential utility of keeping asleep and waiting for the subsequent devices $(l>S^l_i)$ turning on.
Specifically, in the forth segment, we have $L$ components showing the decrease of joint error if the subsequent devices turn on at this time slot, we have additional $L$ components showing the weighted (multiplied by $PD$) decrease of joint error if the subsequent devices turn on, and another constant component $1$ just to indicate the action of $a_0$.
At last, the fifth segment only includes the remaining time of the current state, $T-S^t_i$.
Each value is normalized to one before training $\mathcal{NN}$.

\begin{algorithm}[!thp]
\caption{Deep Q-learning for value approximation in multi-device power control.}\label{alg_Multiple}
\KwIn{Feature vector for any state-action pair $\vec{f}(s,a)$.}
\KwOut{Action strategy $\{\pi(s)=\arg\max_a\tilde{Q}(s,a)\}$.}
\Begin
{
Use a random strategy to generate a training set $\big\{\big<\vec{f}(s,a),Q(s,a)\big>\big\}$ to generate an initial $\mathcal{NN}$\;
Initialize a training data set $\mathcal{D}$\;
\For{episode = $1$ to $N$}
{
    Randomly generate an initial state $S_1$\;
    \While{State transition is not complete}
    {
        The current state is $S_i$\;
        With probability $\epsilon$ choose a random action\;
        With $1\!-\!\epsilon$ select $a=\arg\max_a \tilde{Q}(S_i,a)$\;
        Perform the action, take the reward $R_{S_i}^a$, and the new state is $S_{i\!+\!1}$\;
        \eIf{$S_{i\!+\!1}$ is the termination state}
        {
            Add $\big<\vec{f}(S_i,a),\,R_{S_i}^a\!+\!\max_{a^\prime}\!\tilde{Q}(S_{i\!+\!1},a^\prime)\big>$ into $\mathcal{D}$\;
        }
        {
            Add $\big<\vec{f}(S_i,a),\,R_{S_i}^a\big>$ into $\mathcal{D}$\;
        }
    }
    Randomly select a batch of $B$ training data from $\mathcal{D}$ to perform gradient descent to improve the $\mathcal{NN}$.
}
}
\end{algorithm}

\textbf{Training of the NN:} Massive amount of training data should be provided to obtain an accurate approximation.
We first use a random power control strategy to experience through the state transition procedure.
In this way, a set of $\big<\vec{f}(s,a),Q(s,a)\big>$ values can be collected.
Then we perform a training process to minimize $\sum\big[\mathcal{NN}\big(\vec{f}(s,a)\big) -Q(s,a) \big]^2$ and get the initial training $\mathcal{NN}$ model.
Next, we use the $\mathcal{NN}$ model to perform action decision in a new system, with probability $0\!<\!1-\epsilon\!<\!1$ to select $a=\arg \max \tilde{Q}(s,a)$ for each state (and with $\epsilon$ to randomly select).
In this way, we can create a new episode of the experience and record the tuples of $<\vec{f}(s,a), R_s^a, \tilde{Q}(s^\prime,a^\prime)>$ along the experienced states.
As shown in Eqn.~(\ref{eqn_qrq}), $\sum[\tilde{Q}(s,a) - R_s^a - \tilde{Q}(s^\prime,a^\prime)]^2$ can also be considered as a new minimizing target to improve the current $\mathcal{NN}$.
Therefore, the recorded tuples form a new training set $\big\{\big<\vec{f}(s,a), \,\, R_s^a\!+\!\tilde{Q}(s^\prime,a^\prime)\big>\big\}$.
We can repeatedly use the $\mathcal{NN}$ to perform action decision in new systems and create multiple sets of training data.
To avoid the situation where the $\mathcal{NN}$ may overfit with the newest training data set, we randomly choose a batch of the training data from all the previous training data set and perform limited numbers of gradient descent, just like the way in~\cite{bib_DQN}, which was called as deep Q-learning with experience replay.
An overview of the training procedure is given in Algorithm~\ref{alg_Multiple}.

\section{Location Selection Strategy}\label{sec_LocationSelection}

In this section, we study the problem of location selection based on a fixed power control strategy for $L$ available sensing devices.
Note that the power control $\{\phi_{k,t}\}$ might dynamically change if the environment $\{\mu_t\,|\,t\!=\!0,1,\cdots T\}$ was not fixed.
Therefore, the environment is also considered as a given condition, in which way we can analyze the performance of different location selection schemes.
To ensure that the location selection is not affected by the specific environment setting, $T$ should be set large enough to integrate all possible stochastic situations based on Eqn.~(\ref{eqn_AQIProbTrans}).

The rest of this section describes the solution of selecting $L$ sensing locations from $K$ locations.
To be specific, we first model the correlation of different locations and use a clustering algorithm to create initial sets of $L$ locations.
Then a genetic algorithm is designed to widely search the solution space in order to acquire the best location selection.

\subsection{Correlation of Locations}\label{sec_LocationSelectionCorrelation}

As introduced in Section~\ref{sec_ModelError}, $\mu_{k_1,k_2}$ and $\sigma_{k_1,k_2}^2$ are used to quantitatively describe the statistic relation of the locations $k_1$ and $k_2$.
If we use the value at the location $k_1$ to infer the value at the location $k_2$ when the average air quality is $\mu_t$, then the deviation of the mean value, $\mu_t\!\cdot\!\mu_{k_1,k_2}$, and the additional variance, $\mu_t^2\!\cdot\!\sigma_{k_1,k_2}^2$, should be counted.
A higher value of $|\mu_{k_1,k_2}|$ or a higher value of $\sigma^2_{k_1,k_2}$ indicates that the air quality at $k_1$ and the air quality at $k_2$ have lower similarity, i.e., greater difference.

\textbf{Calculation of $\mu_{k_1,k_2}$ and $\sigma_{k_1,k_2}^2$:} These two parameters are calculated from the collected data set.
We denoted the collected data from location $k_1$ and location $k_2$ as $y(k_1,t)$ and $y(k_2,t)$, with $t=1,2,\cdots,T$.
Therefore, we have:
\begin{align}
& \mu_{k_1,k_2} = \dfrac{1}{T} \sum_{t=1}^T \dfrac{y(k_2,t) - y(k_2,t)}{\mu_t}, \label{eqn_ParaMukk}\\
& \sigma_{k_1,k_2}^2 =  \dfrac{1}{T} \sum_{t=1}^T\dfrac{\big[y(k_1,t) +\mu_t\!\cdot\!\mu_{k_1,k_2} - y(k_2,t)\big]^2}{\mu_t^2}, \label{eqn_ParaSigmakk}
\end{align}
where $\mu_t$, the average air quality value, is acquired based on the data collected from all the locations.

\textbf{Difference Matrix:}
For any two locations, $k_1, k_2 \in{\mathcal{K}}$, we defined $\theta_{k_1,k_2}$ as their difference, which is given by
\begin{equation}\label{eqn_Difference}
\theta_{k_1,k_2} = \sqrt{\mu_{k_1,k_2}^2+\sigma_{k_1,k_2}^2}, \quad k_1, k_2 \in{\mathcal{K}},
\end{equation}
which has a similar form with the definition of the joint error in Eqn.~({\ref{eqn_Joint}}).
We put $\theta_{k_1,k_2}$ in the matrix form, given as below:
\begin{equation}\label{eqn_DifferenceMatrix}
\Theta_{K\times K} = \left[
\begin{array}{cccc}
\theta_{1,1} & \theta_{1,2} & \cdots & \theta_{1,K}\\
\theta_{2,1} & \theta_{2,2} & \cdots & \theta_{2,K} \\
\vdots & \vdots & \ddots & \vdots \\
\theta_{K,1} & \theta_{K,2} & \cdots & \theta_{K,K}
\end{array}
\right].
\end{equation}

\begin{remark}
The difference matrix defined by Eqn.~(\ref{eqn_DifferenceMatrix}) is a symmetric matrix with zeros diagonal elements.
The diagonal elements are zeros because $\mu_{k,k}=0$ and $\sigma_{k,k}^2=0$.
The matrix is symmetric because $\mu_{k_1,k_2}=-\mu_{k_2,k_1}$ and $\sigma_{k_1,k_2}^2=\sigma_{k_2,k_1}^2$, which can be directly deduced from Eqn.~(\ref{eqn_ParaMukk}) and Eqn.~(\ref{eqn_ParaSigmakk}).
\end{remark}

\textbf{High dimensional feature space:} The difference of the locations $k_1$ and $k_2$, $\theta_{k_1,k_2}$, can be considered as the distance of them in a high dimensional space.
This space depicts the implicit features of each location $k \in \mathcal{K}$.
A greater difference (or distance) indicates that the corresponding two locations have less similarity.
Note that to avoid the case like $\theta_{1,3}>\theta_{1,2}+\theta_{2,3}$ (which is not acceptable in Euclidean space), we can add a same small amount value $\delta=\theta_{1,3}-\theta_{1,2}-\theta_{2,3}$ to all the values of $\theta_{k_1,k_2}$, $\forall k_1 \neq k_2$.
In the following, we omit the expression of $\delta$ and consider $\Theta_{K\times K}$ as a valid distance matrix.
We then need to decide the coordinate of each $k \in \mathcal{K}$, denoted by $\vec{x}_k=(x_k^{(1)},x_k^{(2)},\cdots,x_k^{(D)})$, where $D$ is the dimension of the feature space.
Since the difference matrix $\theta_{K\times K}$ does not explicitly express these coordinates, we provide the following method to acquire them.

\textbf{Calculating the coordinates:} First, the dimension $D$ is set to be $D\!=\!K\!-\!1$.
The coordinate of the first location is set to be zero, given by $\vec{x}_1=(0,0,\cdots,0)$.
The coordinate of the second location is $\vec{x}_2=(\theta_{1,2},0,\cdots,0)$.
For the $k^{th}$ coordinate $\vec{x}_k$, we need to calculate the solution of a $(k\!-\!1)$-variable quadratic equation set, given by
\begin{equation}\label{eqn_Coordinate}
\left\{\!\!
\begin{array}{cl}
\big(x_1^{(1)}\!-x_k^{(1)}\big)^2 + \cdots + \big(x_1^{(k-1)}\!-x_k^{(k-1)}\big)^2 = \!\!&\!\! \theta_{1,k}^2 \\
\cdots  \quad\quad\quad\quad\quad\quad\quad\quad\quad\quad \cdots\quad\quad \!\!&\!\! \\
\cdots  \quad\quad\quad\quad\quad\quad\quad\quad\quad\quad \cdots\quad\quad \!\!&\!\! \\
\big(x_{k-1}^{(1)}\!-x_k^{(1)}\big)^2 + \cdots + \big(x_{k-1}^{(k-1)}\!-x_k^{(k-1)}\big)^2 = \!\!&\!\! \theta_{k-1,k}^2 \\
\end{array}
\right. \!\!\!\!\!\!\!\!\!\!\!\!
\end{equation}
where $x_k^{(1)},\cdots, x_k^{(k-1)}$ are the variables to be solved.
In case of multiple solutions, any one can be acceptable since we only need to satisfy the difference matrix $\Theta_{K\times K}$ as the constraint.

\textbf{Clustering:} For any two of the locations, $k_1, k_2 \in \mathcal{K}$, if they have short distance in the feature space, then the joint error of using one to infer another will be low.
Otherwise, the joint error will be relatively greater.
From this intuition, the $L$ locations that we aim to select from $\mathcal{K}$ should be ``as separated as possible".
Therefore, we use the classical $k$-means algorithm~\cite{bib_Kmeans} to cluster the $K$ locations in the feature space into $L$ clusters.
The major weakness of the $k$-means algorithm is the necessity of predefining the value of $k$, which in fact becomes an advantage in our usage since this parameter happens to be fixed as $L$.

\textbf{Initial sets:} With $L$ clusters, we can randomly choose one location from each cluster to create a location set set $\mathcal{L}$, with $|\mathcal{L}|=L$.
Since there could be many possible $\mathcal{L}$, we denote the collection of the location sets as
\begin{equation}
\left\{
\begin{array}{ll}
\mathcal{C}  = \big\{  \mathcal{L}_c,  \big\}, \quad\quad \!\!&\!\! \forall c=1,2,\cdots C, \\
\mathcal{L}_c  = \big\{k_l\big\}, \quad\quad  \!\!&\!\! \forall k_l \in \mathcal{K}  \,\,,\forall l=1,2,\cdots L,
\end{array}
\right.
\end{equation}
where we confine the size of collection as $C$.
The collection $\mathcal{C}$ therefore becomes the initial choices of the location selection, which can be used in the following genetic algorithm.

\subsection{Exploring the Solution Space by Genetic Algorithm}\label{sec_LocationSelectionGenetic}

The reason we use genetic algorithm is because the solution space of this $0$-$1$ integer optimization problem is highly complicated.
Any approximated solution could fall into a minimum value with poor performance.
The genetic algorithm can widely search the solution space and provide a satisfying outcome~\cite{bib_Genetic}.
In the following, we introduce our implemented genetic algorithm, including genetic coding, genetic recombination, genetic mutation, genetic selection and the termination criteria.

\textbf{Genetic coding:} For each location set $\mathcal{L}$, we use a $K$-length vector to indicates the corresponding coded gene.
A gene of the location set $\mathcal{L}$ is denoted as
\begin{equation}
\vec{G}(\mathcal{L})=\big(G^{(1)}(\mathcal{L}), G^{(2)}(\mathcal{L}), \cdots, G^{(K)}(\mathcal{L})\big),
\end{equation}
where $G^{(k)}(\mathcal{L})$ is a boolean function, indicating whether the $k^{th}$ location is included in the location set $\mathcal{L}$.
For each valid location set (i.e., up to $L$ selected locations), the number of ``1"s in its coded gene is no more than $L$.
For the initial collection of location sets $\mathcal{C}$ introduced in Section~\ref{sec_LocationSelectionCorrelation}, we encode them and add their genes to a gene pool, denoted by $\mathcal{G}$.

\textbf{Genetic mutation:} The mutation of a gene is to create the possibility to randomly try another solution.
Each gene creates $M$ copies of itself, where each bit of the gene has a fixed possibility $p_m$ to mutate during the copy ($0\!\rightarrow\!1$, or $1\!\rightarrow\!0$).
Therefore, different genes (solutions) are generated and added to the gene pool $\mathcal{G}$.
Note that the number of $1$ may not satisfy the constraint, which will be considered in the following genetic selection part.

\textbf{Genetic recombination:} For any two existing genes, a recombination procedure can be performed.
To be specific, for $\vec{G}_1$ and $\vec{G}_2$, we randomly choose a position $g=1,2,\cdots K-1$ and exchange the bits of these two genes behind the position~$g$.
The new genes are $\vec{G}_1^\prime=\big(G_1^{(1)}, \cdots, G_1^{(g)}, G_2^{(g+1)}, \cdots , G_2^{(K)}\big)$ and $\vec{G}_2^\prime=\big(G_2^{(1)}, \cdots, G_2^{(g)}, G_1^{(g+1)}, \cdots , G_1^{(K)}\big)$.
These new genes are also added into the gene pool $\mathcal{G}$.

\begin{figure}[!thp]
\centering
\includegraphics[width=3.3in]{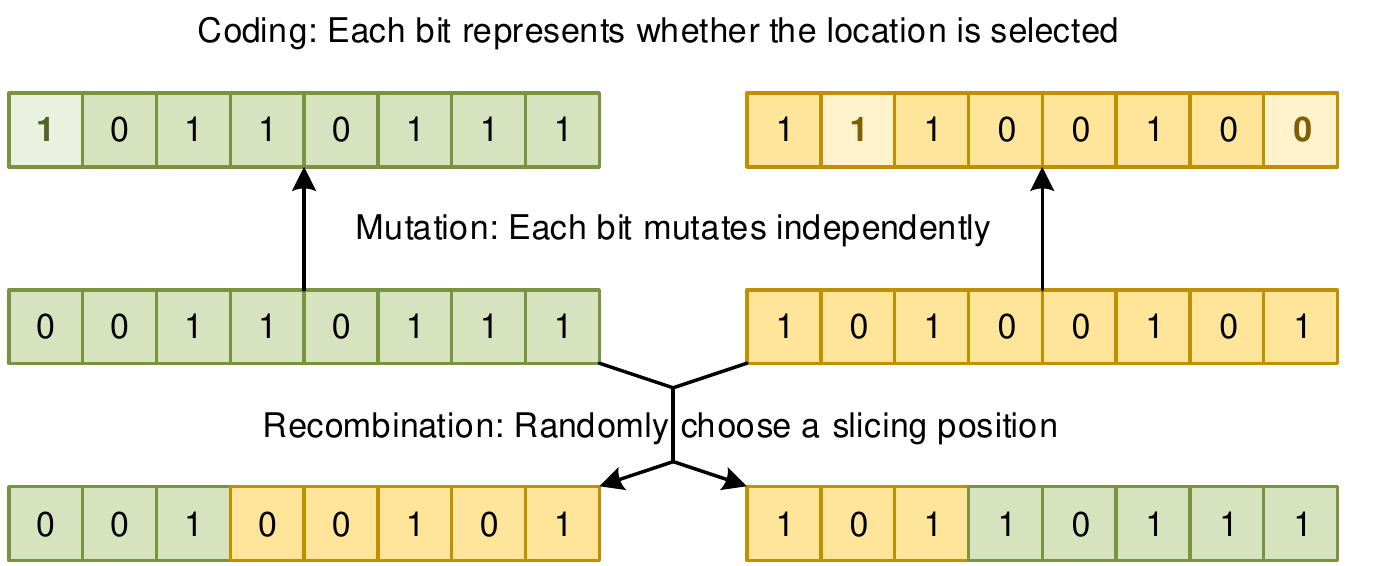}
\vspace{-2mm}
\caption{An example of genetic coding, mutation and recombination.}\label{fig_Genetic}
\vspace{-4mm}
\end{figure}

\textbf{Genetic selection:} The size of the gene pool $\mathcal{G}$ is not unlimited.
As the new genes generated by mutation and recombination are added into the gene pool, there could be too many genes.
A selection procedure should be performed to choose the good ones and eliminate the bad ones.
First, the duplicated genes and the genes with more than $L$ ``1"s are eliminated.
Then we check the disadvantage of each existing gene, which is defined as $\bar{J}$ of the corresponding location set based on the given power control and environment.
Note that the aforementioned power control matrix $\Phi_{K\times T}$ has $L\!<\!K$ non-zero vectors, which indicates the power control for the given $L$ locations.
Here, we need to interpret $\Phi_{K\times T}$ as $\Phi_{L\times T}$, with $L$ power control schemes for each location in the location set.
If the number of ``1"s in a gene is fewer than $L^\prime \!<\!L$, then we only choose first $L^\prime$ vectors as the corresponding power control schemes.
With the disadvantage of each existing gene, we keep the best $H_1$ ones in the gene pool, and then randomly choose $H_2$ genes by using the disadvantage of each gene as its weighted probability.
Such probability is given by
\begin{equation}\label{eqn_Selection}
p_g \propto \Big[\Big(\max\limits_{g}\bar{J}\Big) - \bar{J}_g \Big] , \quad\quad \forall g \in \mathcal{G}.
\end{equation}
Therefore, the gene pool only keeps $H\!=\!H_1+H_2$ genes after each round of genetic selection.

\begin{algorithm}[!thp]
\caption{Genetic Algorithm for Location Selection.}\label{alg_Location}
\KwIn{The power control strategy $\{\phi_{k,t}\}$ and the environment condition $\{\mu_t\}$.}
\KwOut{Optimal location set $\{L\}$.}
\Begin
{
Calculate the coordinates of the locations in the feature space according to~(\ref{eqn_Coordinate})\;
Use $L$-means algorithm to perform clustering and acquire $L$ clusters\;
Randomly choose one location in each one of the cluster to create a location set\;
Generate $C$ location sets and put their coded genes into the gene pool $\mathcal{G}$\;
\For{Evolution count $w=1$ to $W$}
{
    Perform genetic mutation and recombination\;
    Eliminate the duplicated genes and the genes with too many ``1"s\;
    Calculate the disadvantage of each existing gene\;
    Select the best $H_1$ genes and randomly select $H_2$ genes according to Eqn.~(\ref{eqn_Selection})\;
    \If{no improvement for $6$ iterations}
    { Break the evolution\; }
}
Choose the gene with the lowest disadvantage\;
}
\end{algorithm}

\textbf{Evolution and termination:} For each round of genetic evolution, based on the current gene pool, we perform the genetic mutation for each existing gene and the genetic recombination between randomly chosen gene pairs.
As new genes are added, we then perform genetic selection to keep only $H$ genes and record the lowest value of disadvantage.
If the best performance hasn't been improved for $6$ rounds, or it has been more than $W$ rounds, then the evolution process terminates.
Otherwise, we repeat the mutation and recombination to continue the genetic evolution.
An overview of the genetic algorithm is provided in Algorithm~\ref{alg_Location}.

\section{Emulation}\label{sec_Evaluation}

In this section, we evaluate the performance of the proposed power control strategy and the proposed location selection strategy.
Simulation setups are given in Section~\ref{sec_SimulationSetup}, simulation results and corresponding discussions are provided in Section~\ref{sec_SimulationResult}.

\subsection{Data Set Usage and Parameters setup}\label{sec_SimulationSetup}

Some of the parameters are extracted from our collected data, which is based on $30$ air quality sensing devices deployed in Peking University for around seven months~\cite{bib_DataSet}.
The sensing interval (the length of the time slot) is $10$ minutes, and each collected data is an integer representing the detected PM2.5 value at the given location and time.

We denote the measured value at the $k^{th}$ location at time $t$ as $y(k,t)$.
Therefore, the normalized measurement variance $\sigma_0^2$ and the temporal deviation variance $\sigma_{d}^2$ is given by:
\vspace{-1mm}
\begin{align}
& \sigma_0^2 = \dfrac{1}{T} \sum_{t=1}^T \dfrac{1}{L} \sum_{k \in \mathcal{K}_L}\dfrac{[y(k,t)-\mu_t]^2}{\mu_t^2}, \\
& \sigma_{d}^2 = \dfrac{1}{T\!-\!1} \sum_{t=1}^{T-1} \dfrac{1}{L} \sum_{k \in \mathcal{K}_L} [y(k,t\!+\!1)-y(k,t)]^2 .
\end{align}
For the parameters describing the relation of the nodes, $\{ \mu_{k_1,k_2} \}$ and $\{ \sigma_{k_1,k_2}^2 \}$, we have already provided them in Eqn.~(\ref{eqn_ParaMukk}) and Eqn.~(\ref{eqn_ParaSigmakk}).

The deep neural network in Section~\ref{sec_PowerControlMultiple} is designed to have $5$ hidden layers.
Given the location number $K$ and device number $L$, each hidden layer is designed to have $4K\!+\!L, 4K, 3K, 2K,$ and $K$ neurons, respectively (from the input side to the output side).
Each round of gradient descend of the neural network is realized by running the ``trainscg" algorithm provided in MATLAB for one epoch on the training data set.

A more detailed parameter setting is listed in Table~\ref{tab_Parameters}.

\begin{table}[!thp]
\renewcommand\arraystretch{1.2}
\caption{Simulation parameters}\label{tab_Parameters} \centering
\begin{tabular}{|p{52mm}|p{30mm}|}
\hline
Normalized measurement variance, $\sigma_0^2$ & $0.0037$ \\
\hline
Temporal deviation variance, $\sigma_{d}^2$ & $10.89$ \\
\hline
Normalized mean relation, $\{ \mu_{k,k^\prime} | {k\!\neq\!k^\prime}\}$ & between $-0.15$ and $0.15$ \\
\hline
Normalized variance relation, $\{ \sigma_{k,k^\prime}^2 | {k\!\neq\!k^\prime} \}$ & between $0.001$ and $0.1$ \\
\hline
Air quality values, $\{\mu_t\}$ & between $1$ and $508$ \\
\hline
Total number of time slots, $T$ & between $500$ and $10000$ \\
\hline
Total number of locations, $K$ & $30$ \\
\hline
Number of available devices, $L$ & between $5$ and $25$ \\
\hline
Energy budget of each device, $E$ & between $100$ and $1000$ \\
\hline
Maximum allowable sleep time slots, $\Delta T$ & between $10$ and $12$ \\
\hline
Episode number to train $\mathcal{NN}$, $N$ & between $1$ and $100$ \\
\hline
Batch size to improve $\mathcal{NN}$ in each round, $B$ &$100000$ \\
\hline
Random action probability, $\epsilon$ & from $0.1$ to $0$ \\
\hline
Maximum number of rounds of evolution, $W$ &$25$ \\
\hline
Size of gene pool, $H$ and $C \, (H=C)$ & between $40$ and $100$ \\
\hline
Two types of genetic selection, $H_1$ and $H_2$, & $0.1H$ and $0.9H$ \\
\hline
Copy number during genetic mutation, $M$ &$3$ \\
\hline
Mutation probability for each bit, $p_m$ & $0.1$ \\
\hline
\end{tabular}
\end{table}

\subsection{Simulation Results and Discussions}\label{sec_SimulationResult}

In this subsection, we first testify the feasibility of the proposed Gaussian reference model based on statistical validation.
Then we take a look at the result of the single device power control plan.
After that, we observe the performance of the multi-device power control and the outcome of the location selection strategy.
And finally, the combination of multi-device power control and location selection is presented to show the joint advantage.

\begin{figure}[!thp]
\centering
\includegraphics[width=3.5in]{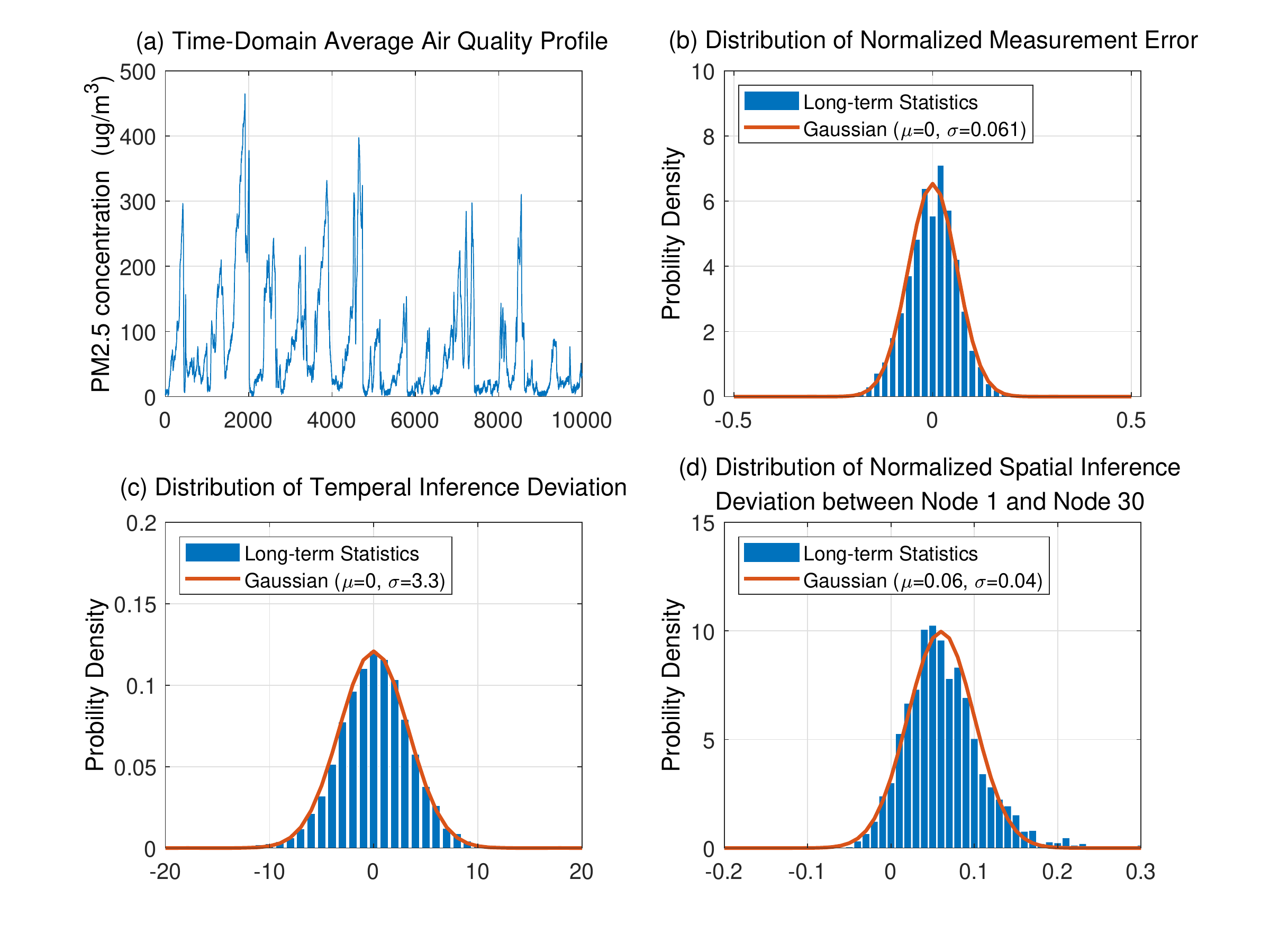}
\vspace{-3mm}
\caption{The testification results of the comparison between the collected data and the Gaussian distribution.}\label{fig_CoverLetterModel}
\vspace{-3mm}
\end{figure}

\textbf{Model Validation:} We first use the collected data to testify the feasibility of using Gaussian model to perform inferences.
As shown in Fig.~\ref{fig_CoverLetterModel}, we provide four subplots to show the statistics of our data.
The subplot (a) illustrates a time-domain profile of the average air quality (of 30 devices) with 10000 continuous time slots (from March 1st to May 15th).
A couple of waves of ``haze weather" with different peak values can be observed.
In the subplot (b), we provide the distribution of the normalized measurement error of the sensors.
A zero-mean Gaussian distribution profile with $\sigma=0.061$ is used to fit the statistics.
The subplot (c) shows the distribution of temporal inference deviation.
It can be observed that it is quite accurate to use Gaussian distribution to model it.
Finally, in subplot (d), the distribution of normalized spatial inference deviation is provided\footnote{Here we have to exclude the time slots with small values of $\mu_t$ due to the following reason: The values of the collected data are an integers (such as PM2.5 values), which are not continuous. If $\mu_t$ is low, the difference of two locations mostly resides in $\{-1,0,1\}$. The resolution of its distribution is low and the final distribution will have a peak at the location of $x\!=\!0$. This is actually the problem of quantization accuracy, which is not the focus of our paper. Therefore, we only calculate the time slots with $\mu_t>30$ to get the distribution in the subplot (d).}.
Without the loss of generality, we only provide the difference between of Node 1 and Node 30 for illustration.
The corresponding similarity with Gaussian distribution is not as good as (b) and (c).
However, it is not a vital problem since we only aim to provide a rough and simple inference method but not an accurate but unanalyzable method as discussed in Section~\ref{sec_ModelError}.

\begin{figure}[!thp]
\centering
\includegraphics[width=3.3in]{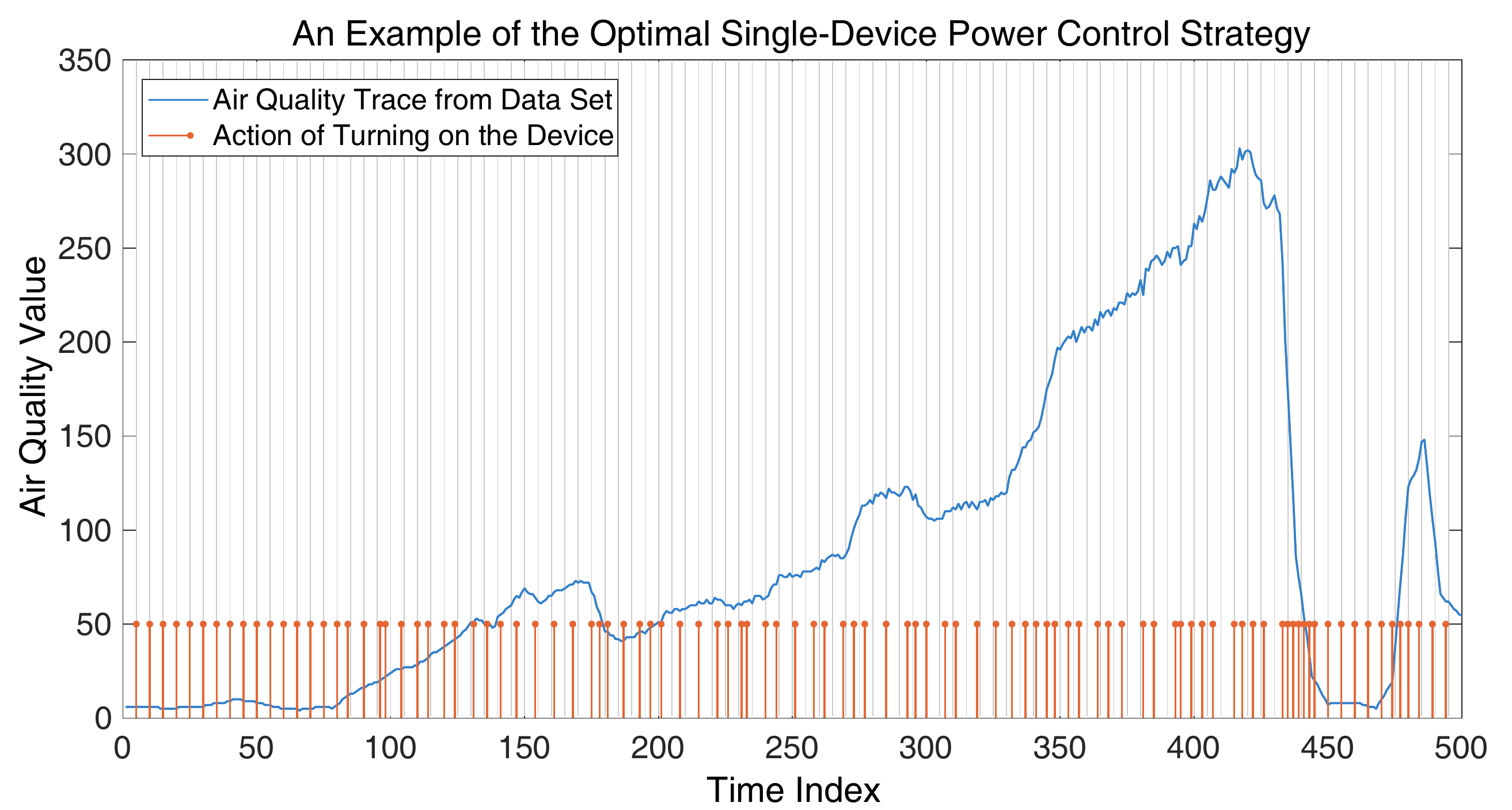}
\vspace{-1mm}
\caption{An example of the optimal single-device control strategy, with $T=500$, $E=100$, and $\Delta T=12$. The sensing actions are performed on the times of those orange sticks.}\label{fig_SimulationSingle}
\vspace{-3mm}
\end{figure}

\textbf{Single-device power control:} We then illustrate the characteristic of the optimal solution for the single-device power control.
The result is presented in Fig.~\ref{fig_SimulationSingle}, where we set the number of time slots $T\!=\!500$, the energy budget $E\!=\!100$, and the maximum allowable sleep time $\Delta T\!=\!10$.
The grey vertical lines represent the uniform sensing strategy, which has $T/E\!=\!5$ time slots between adjacent sensing actions.
The optimal sensing actions are presented in the form of those short orange lines, which are not uniformly distributed.
It can be observed that the sensing actions are more frequent if the air quality changes rapidly.
This is because the value of $|\mu-\mu_t|$ will significantly increase if we rely on temporal inference.
In this figure, the average joint error $\bar{J}$ of the uniform sensing is $11.05$, while the optimal power control only leads to $\bar{J}\!=\!10.43$ indicating an around $6\%$ performance gain.

\begin{figure}[!thp]
\centering
\includegraphics[width=3.5in]{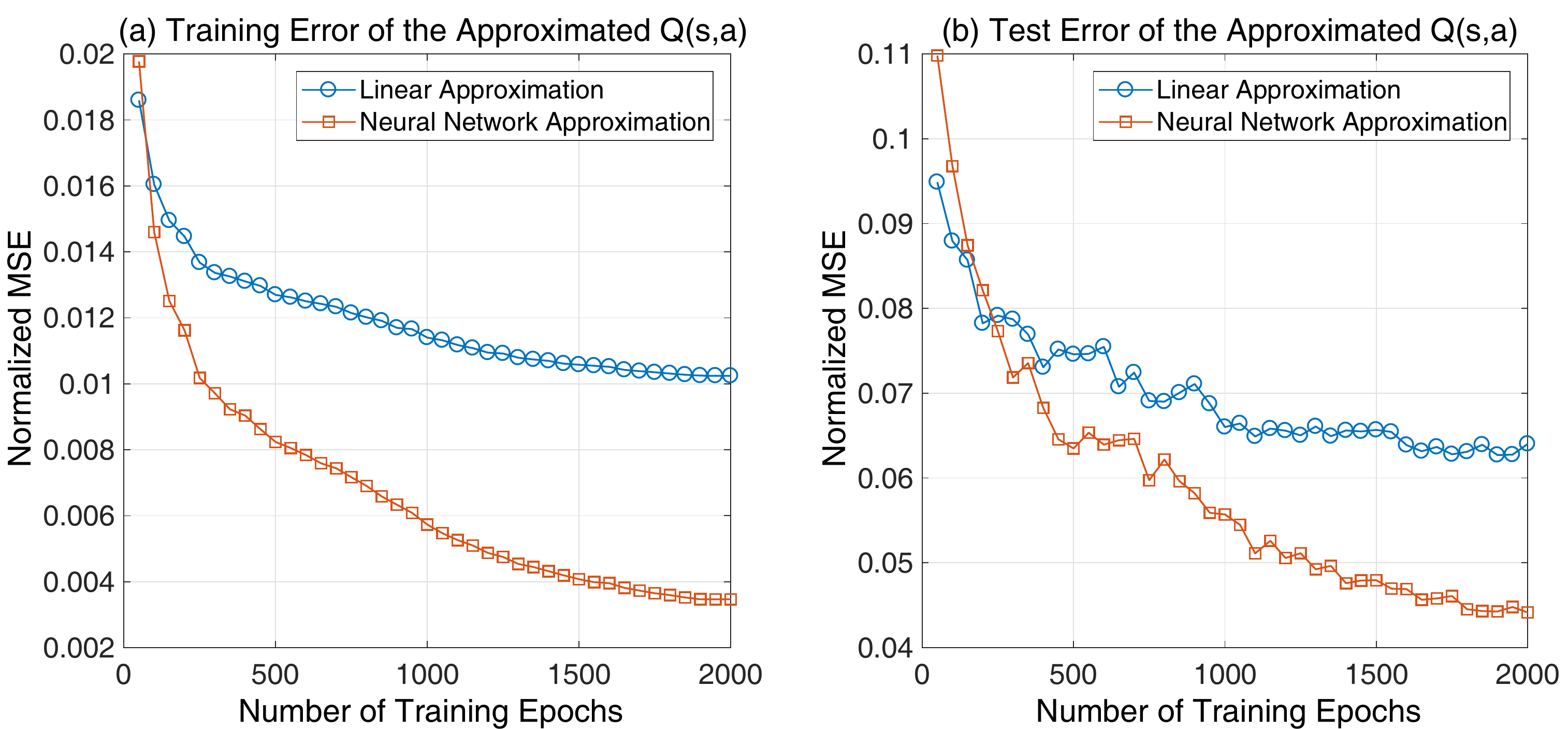}
\vspace{-3mm}
\caption{The training error and test error of the approximated Q(s,a) function. The subplot (a) shows the normalized MSE during training. The subplot (b) shows the normalized MSE when perform a test on another data sets.}\label{fig_SimulationTrainError}
\vspace{-3mm}
\end{figure}

\textbf{Multi-device power control:} The multi-device power control involves the deep Q-learning which approximates the $Q(s,a)$ function.
Before providing its performance, we first observe the accuracy of such approximation, as given in Fig.~\ref{fig_SimulationTrainError}, where we set $T\!=\!10000$, $E\!=\!2000$, $\Delta T \!=\! 12$, $K\!=\!30$, and $L\!=\!20$.
The subplot (a) shows the change of Mean Square Error (MSE) when we generate the initial neural network.
The accuracy of the neural network is compared with a linear approximation function using the same feature vector as discussed in Section~\ref{sec_PowerControlMultiple}.
It can be seen that the neural network has a better fitting result since the minimum value of MSE is be much lower than that of the linear approximation.
The subplot (b) shows the test error of the trained neural network by using an additional data set, which is not included in the training data set.
It can be seen that the test error is much greater than the training error for both linear approximation and the neural network.
However, there is still a diminishing trend of the test error as we keep training the neural network, which indicates that it is able to use a limited set of training data to approximate other potential data set.

\begin{figure}[!thp]
\centering
\includegraphics[width=3.5in]{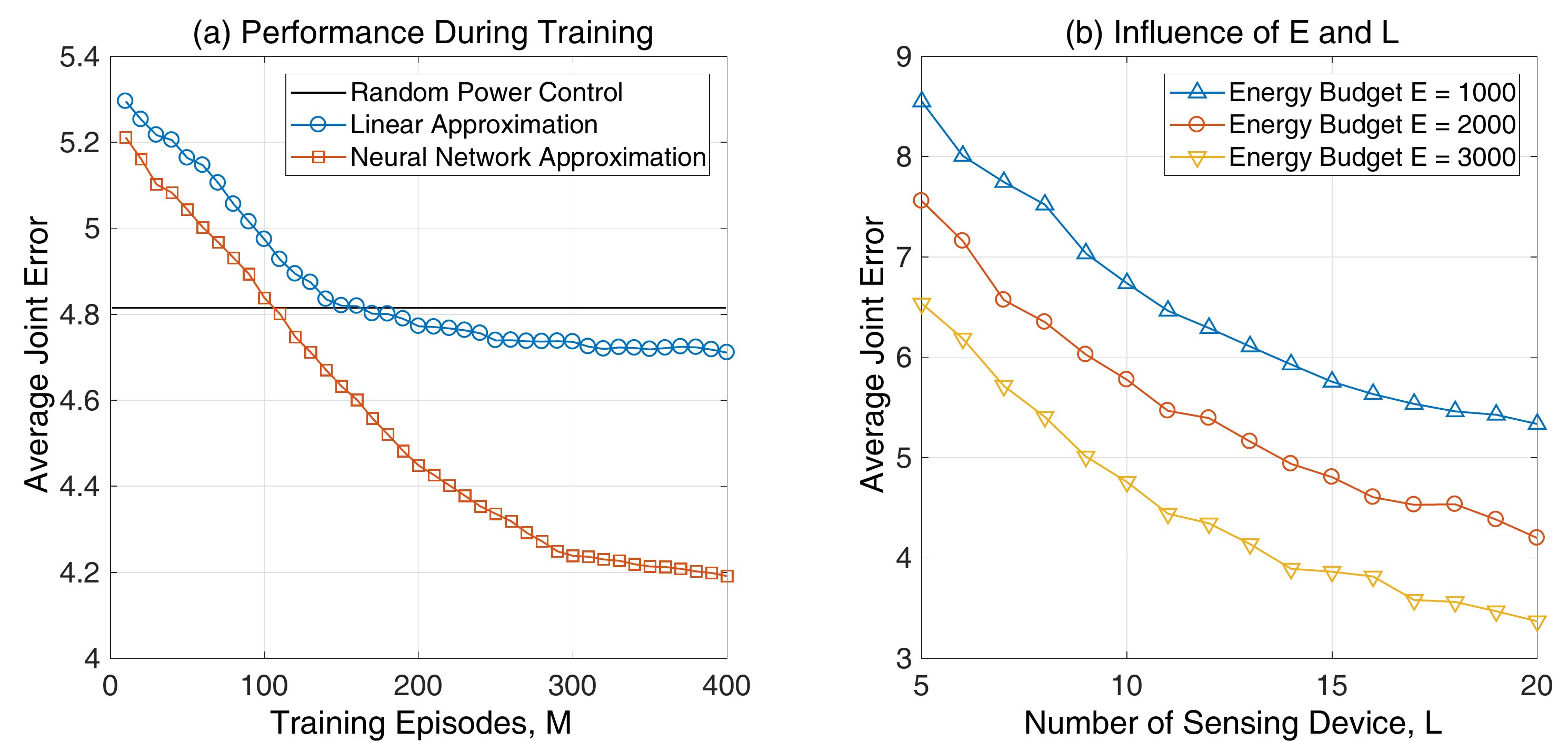}
\vspace{-3mm}
\caption{The performance of the multi-device power control algorithm based on reinforcement learning. The subplot (a) shows performance improvement during the training procedure. The subplot (b) shows the influence of the energy budget and the number of available devices.}\label{fig_SimulationMultiple}
\vspace{-3mm}
\end{figure}

Fig.~\ref{fig_SimulationMultiple} provides the performance of the proposed multi-device power control strategy based on Q-learning.
The subplot (a) shows the performance improvement during the iterative training process, where we set $T\!=\!10000$, $E\!=\!2000$, $\Delta T \!=\! 12$, $K\!=\!30$ and $L=20$.
It can be seen that the linear approximation only has a minor advantage over the random power control scheme, while the proposed deep Q-learning scheme shows a much promising performance.
The subplot (b) presents the influence of the number of sensing devices and the energy budget of the sensing devices, where $T\!=\!10000$, $\Delta T \!=\! 12$, and $K\!=\!30$.
There is a negative correlation of the average joint error $\bar{J}$ and the number of devices $L$, which testifies Proposition~\ref{pro_DeviceNumber}.
And there is a negative correlation of the average joint error $\bar{J}$ and the energy budget $E$, which testifies Proposition~\ref{pro_Power}.
It can also be observed that a greater number of devices would leads to a smaller marginal utility to decrease the joint error.

\begin{figure}[!thp]
\centering
\includegraphics[width=3.5in]{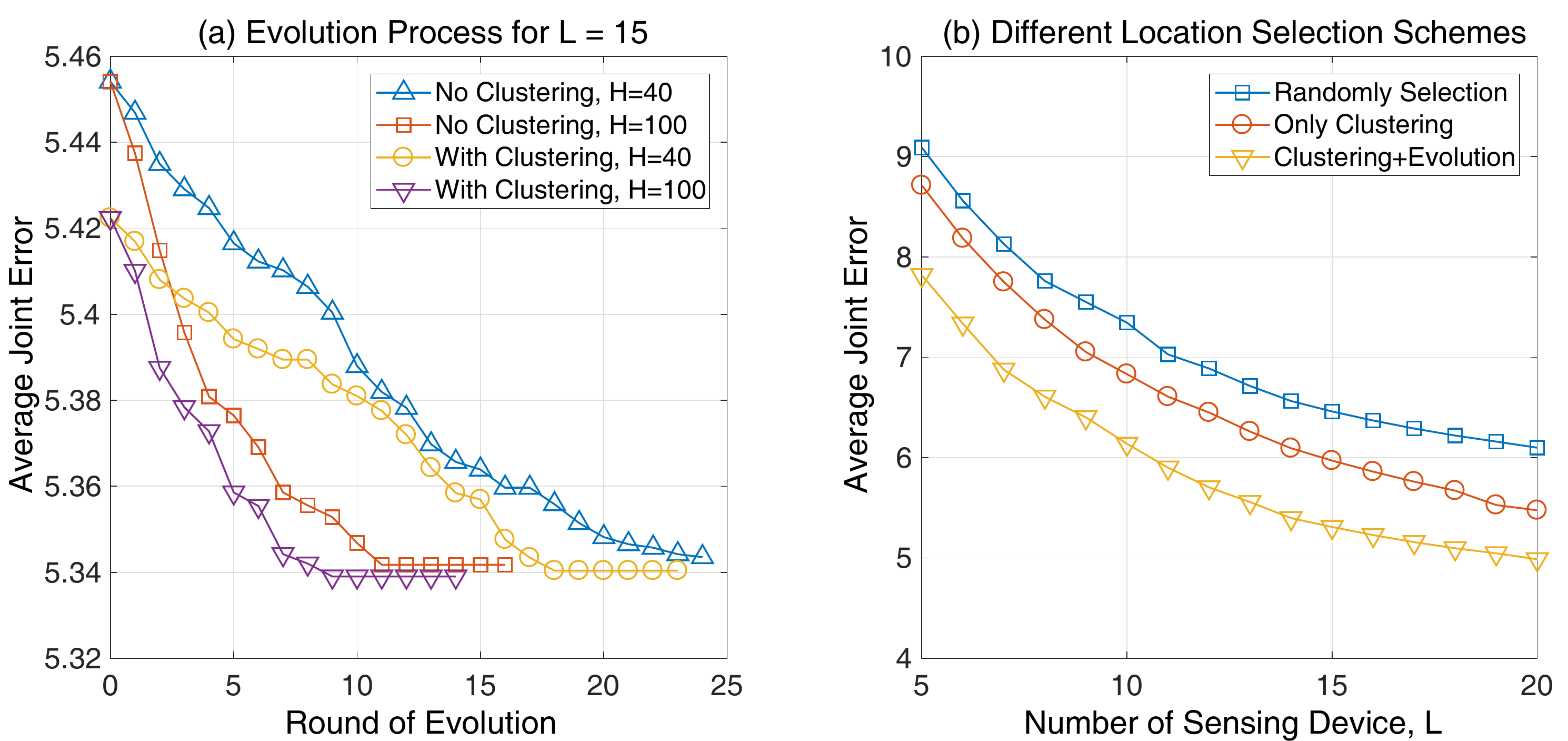}
\vspace{-1mm}
\caption{The outcome of the proposed genetic evolution scheme for location selection. The subplot (a) shows the results of the genetic algorithm with/without initial clustering and different size of gene pool $H$. The subplot (b) shows the influence of the number of available devices $L$ on different schemes.}\label{fig_SimulationLocationSelection}
\vspace{-3mm}
\end{figure}

\textbf{Location selection:} In Fig.~\ref{fig_SimulationLocationSelection}, we present the location selection strategy based on the proposed genetic algorithm, where we set the mutation probability $p_m\!=\!0.1$, copy number $M\!=\!3$ and the maximum number of evolution rounds $W\!=\!25$.
The subplot (a) shows the improvement of performance during the evolution.
Specifically, we have four different settings for comparison by considering the size limit of the gene pool and whether use the clustering to initialize the gene pool.
It can be seen that by using the clustering algorithm, the starting point of the system has a better performance and therefore leads to a shorter convergence time.
In addition, the $100$-sized gene pool has an apparently quicker evolution speed compared with the $40$-sized gene pool, at the cost of a higher memory occupation and a greater amount of computation in each round.
The four of these lines do not converge to the same value of $\bar{J}$, implying that the solution is still sub-optimal.

\begin{figure}[!thp]
\centering
\includegraphics[width=2.5in]{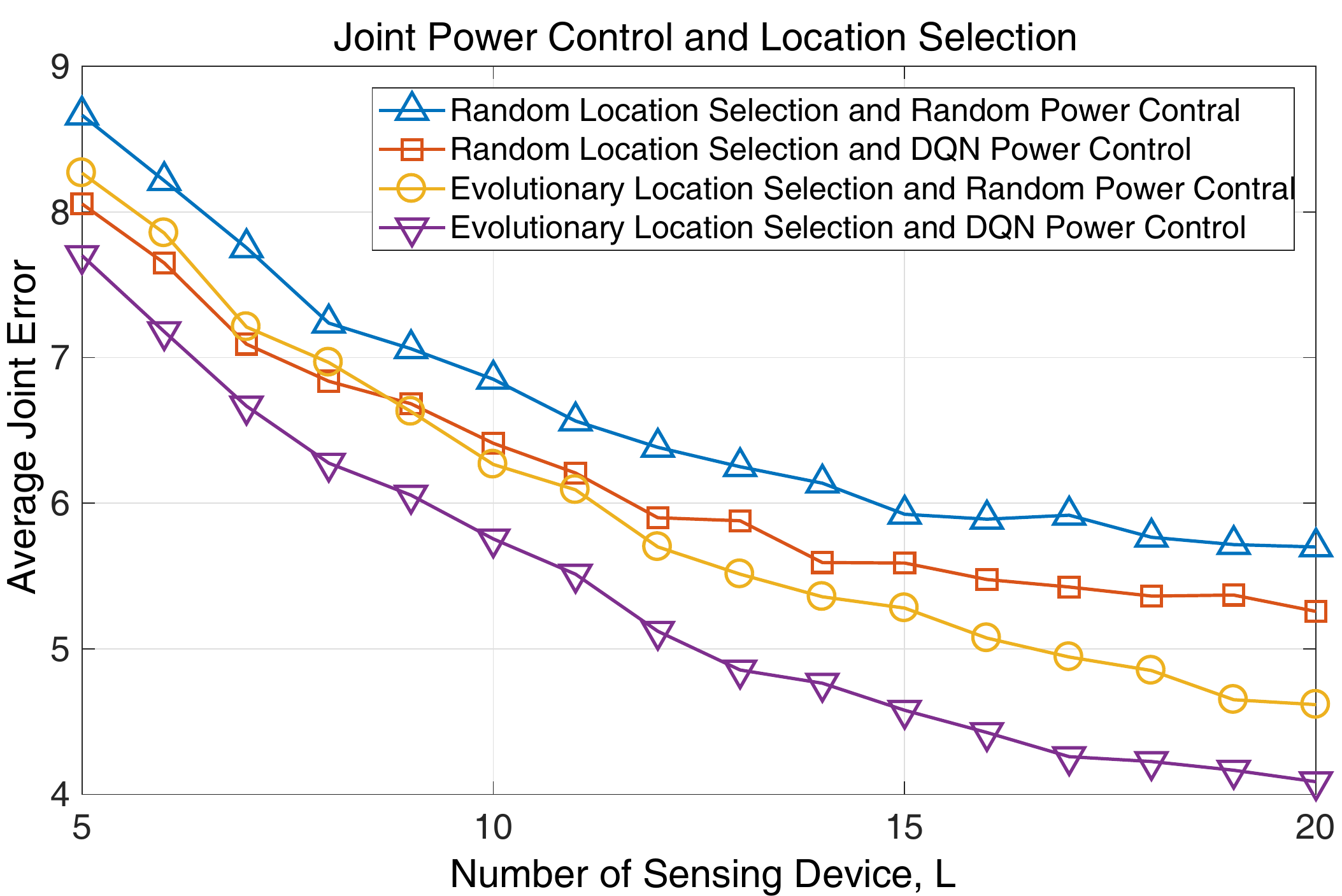}
\vspace{-3mm}
\caption{The performance showing the result of the combination of location selection and power control.}\label{fig_SimulationJoint}
\vspace{-3mm}
\end{figure}

\textbf{Combination of power control and location selection:} Finally, we jointly consider the location selection and the power control strategies, as presented in Fig.~\ref{fig_SimulationJoint}, with $T\!=\!10000$, $\Delta T\!=\!12$, $E\!=\!2000$, $H\!=\!100$, and $K\!=\!30$.
The results of four different combinations of the schemes are included, by considering using the two different power control schemes and two different location selection schemes.
The upmost curve that represents the random location selection and random power control has the highest average joint error $\bar{J}$.
For a specific power control strategy, the evolutionary location selection outperforms the random location selection.
And for a specific location selection strategy, the deep Q-learning power control outperforms the random power control.
It is also noticeable that the evolutionary location selection has a more obvious gain compared to the deep Q-learning power control when the value of $L$ is high, which causes the crossover of the second and the third curve in Fig.~\ref{fig_SimulationJoint}.

\section{Conclusion}\label{sec_Conclusion}

In this paper, we propose the architecture, implementation and optimization of our own air quality sensing system, which provides real-time and fine-grained air quality map of the monitored area.
Specifically for the optimization, we studied the problem of power control and location selection of the air quality sensing system in a smart city.
Our objective was to minimize the joint error of the real-time and fine-grained air quality map, which involved inaccurate data inferences.
We first studied the problem of power control in a stochastic environment based on a fixed location selection and then we studied the problem of location selection based on a fixed power control strategy in a given environment.
The proposed power control strategy was based on deep Q-learning by re-modeling the problem as a MDP.
And the proposed location selection strategy was based on genetic evolutionary algorithm which widely search the solution space.
To evaluate the proposed solution, we extracted the properties from our data set based on our own air quality sensing system deployed in Peking University.
The simulation result showed that the proposed deep Q-learning power control strategy provided a satisfying performance after learning $200$ episodes.
And the proposed genetic evolutionary location selection could quickly achieve a suboptimal solution only by using a small gene pool with the size of $100$.


\end{document}